\documentclass[twocolumn,10pt]{article}

\usepackage{tikz}
\usepackage{amsmath}

\usepackage{filecontents}
\usepackage[english]{babel}
\usepackage{blindtext}

\usepackage{tikz}
\usepackage{amsmath}

\usepackage[english]{babel}
\usepackage{blindtext}
\usepackage{times}
\usepackage{microtype}
\usepackage{url}

\usepackage{graphicx}
\usepackage{tabularx}
\usepackage{subfigure}
\usepackage{booktabs}
\usepackage{siunitx}
\usepackage{xspace}
\usepackage[linesnumbered,ruled]{algorithm2e}
\usepackage{color,soul}
\usepackage{multirow}
\usepackage{tcolorbox}
\newcommand{\tonic}[0]{\texttt{N3IC}\xspace}
\newcommand{\nnp}[0]{\texttt{NNtoP4}\xspace}
\newcommand{\bnne}[0]{\texttt{bnn-exec}\xspace}

\begin{document}

\date{}

\title{\Large \bf Running Neural Network Inference on the NIC}


\author{
{\rm Giuseppe Siracusano}\\
NEC Laboratories Europe
\and
{\rm Salvator Galea}\\
University of Cambridge
\and
{\rm Davide Sanvito}\\
NEC Laboratories Europe
\and
{\rm Mohammad Malekzadeh}\\
Imperial College London
\and
{\rm Hamed Haddadi}\\
Imperial College London
\and
{\rm Gianni Antichi}\\
Queen Mary University of London
\and
{\rm Roberto Bifulco}\\
NEC Laboratories Europe
}


\maketitle

\begin{abstract}
%
%
%
%
In this paper we show that the data plane of commodity programmable Network Interface 
Cards (NICs) can run neural network inference tasks required by packet monitoring 
applications,  with low overhead. This is particularly important as the data transfer 
costs to the host system and dedicated machine learning accelerators, e.g., GPUs, 
can be more expensive than the processing task itself. We design and implement our 
system -– \tonic –- on two different NICs and we show that it can greatly benefit 
three different network monitoring use cases that require machine learning inference 
as first-class-primitive. \tonic can perform inference for millions of network flows 
per second, while forwarding traffic at 40Gb/s. Compared to an equivalent solution 
implemented on a general purpose CPU, \tonic can provide 100x lower processing latency, 
with 1.5x increase in throughput.
%
\end{abstract}

\section{Introduction}
\label{sec:intro}
With the slowdown of Moore's law~\cite{theRiseOfDarkSilicon,esmaeilzadeh11,hardavellas11}, 
the architecture of modern servers is evolving to include an increasing number of domain-specific 
co-processors dedicated to tasks such as cryptography, video processing and machine learning~\cite{hennessy19}. 
Network Interface Cards (NICs) are also following the same trend and now integrate programmble 
components, i.e., Systems-on-Chip (SoC), FPGA, within their hardware data path~\cite{firestone18,netronome_edge,luo18,gupta15}
to lower the pressure on the host CPU, which is usually the bottleneck~\cite{ousterhout15}.
Several efforts have already took advantage of end-host hardware programmability to improve data 
center scalability~\cite{firestone17,firestone18}, applications performance~\cite{eran19} or 
specific functions such as load balancing~\cite{ballani15}, consensus protocols~\cite{kim18} and 
key-value store~\cite{li17}. Lately, an increasing number of networking-related applications 
require machine learning inference as first-class-primitive~\cite{feamster18,liang19}. However, 
the complex operations required by machine learning algorithms, specially deep neural networks,  
often do not allow such applications to effectively leverage programmable NICs~\cite{geng19}.

\begin{figure}[t!]
    \centering
    \includegraphics[width=1\columnwidth]{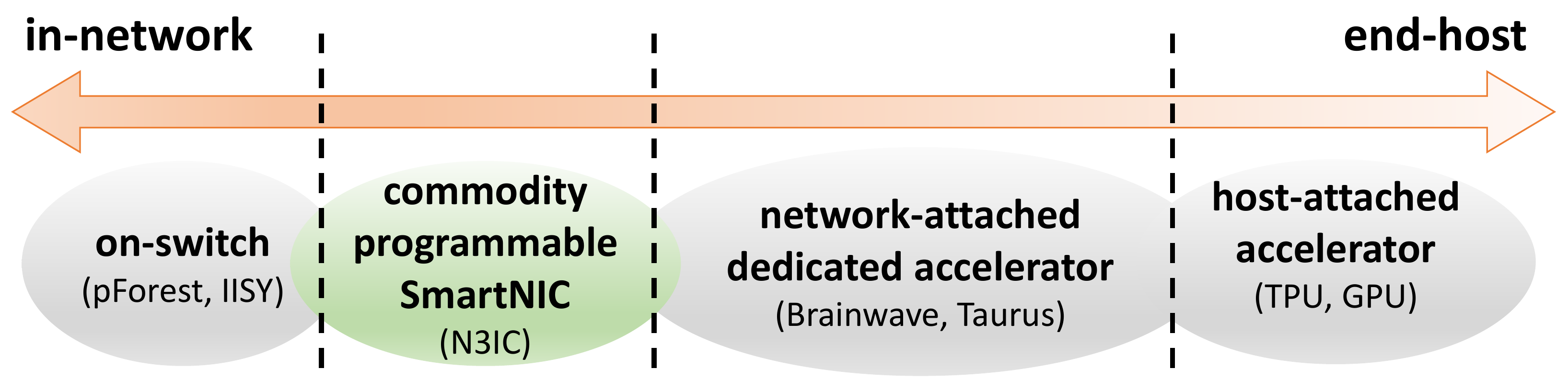}
    \vspace{-0.8cm}
    \caption{Design space for offloading machine learning inference tasks from the end-host's CPU.}
    \label{fig:design-space}
    \vspace{-0.3cm}
\end{figure}

In response to this, we design, implement, and evaluate \tonic, a solution for adding a lightweight neural 
network (NN) inference support to the list of tasks that can be efficiently run on commodity NICs. 
This idea aligns with the current trend of research that moves machine learning capabilities into 
programmable hardware, but takes a novel direction (See Figure~\ref{fig:design-space}). Recent 
efforts focused either on implementing basic machine learning algorithms, e.g., decision trees, 
using programmable switches~\cite{busse19,xiong19}, or deploying expensive and dedicated fully-fledged 
network-attached accelerators (e.g., BrainWave~\cite{brainwave} and Taurus~\cite{swamy19}). In \tonic, 
we design and implement a NN executor that mainly uses already available hardware in the data plane of off-the-shelf programmable NICs, instead. 
NNs are general-purpose machine learning algorithms that suite a growing set of applications~\cite{haddadi,kraska18,kraska19}, 
but they are widely considered too resource-hungry and computationally complex to run on commodity NICs~\cite{busse19,xiong19}. 
We demonstrate that modern programmable NICs can run NN inference, thereby offloading a 
significant amount of computation from the CPUs, and enabling applications with strict timing 
constraints. In fact, we show that PCI Express (PCIe) data transfer overheads~\cite{neugebauer18} 
to the host system can be more expensive than the actual NN inference process (\S\ref{sec:motivation}).

We share the challenges of implementing NN executors taking into account both resource and computational 
constraints of today's commodity programmable NICs (\S\ref{sec:solution}). We then report the 
design of \tonic (\S\ref{sec:impl}) on two different hardware targets: a System-on-Chip (SoC) based 
NIC from Netronome~\cite{netronome} , and an FPGA-accelerated NIC (NetFPGA~\cite{netsume}). Our 
solution relies on a quantization technique known as binarization~\cite{trainingBNN}, which reduces 
the number of bits used to represent NN's inputs and parameters. The resulting NN, i.e., Binary 
Neural Network (BNN), has low memory requirements in the order of KBytes, and its computation can 
be performed using much lighter mathematical operations with small impact on the considered use 
cases' inference accuracy (\S\ref{sec:use_cases}). We provide three implementations of \tonic. 
First, leveraging existing NIC programming languages' primitives, using microC for the Netronome NIC 
and P4 for the NetFPGA. In this last case, we developed a compiler that derives fully functional 
P4 code from a given NN architecture, which can be then synthetized in hardware with the P4-NetFPGA 
framework. Finally, we provide a third implementation that realizes NN inference with a dedicated 
hardware circuitry, which can be exposed as a new primitive to the high-level languages. 

To evaluate the versatility of \tonic, we implemented three different use cases related to network
monitoring: (i) traffic classification, (ii) anomaly detection, and (iii) network tomography. In 
the first two, we seek to identify specific patterns into the traffic data. The latter infers occupation 
levels of a datacenter's switch queues using probe packets, by implementing a slightly modified version of SIMON~\cite{geng19}.
Using the presented implementations, we show (\S\ref{sec:evaluation}) that \tonic can greatly improve 
the performance of typical networking applications that need relatively small NNs. Our evaluations show 
that \tonic can perform traffic analysis for millions of network flows per second, while 
forwarding traffic at 40Gb/s. Compared to a similar system implemented entirely on a general purpose 
CPU, \tonic can provide 100x lower processing latency, 1.5x higher throughput, and save precious 
CPU's cores. Furthermore, \tonic allows applications to run NN inference even 
when their time budget is of few \si{\micro\second}.

\begin{figure}[t!]
    \centering
    \includegraphics[width=1\columnwidth]{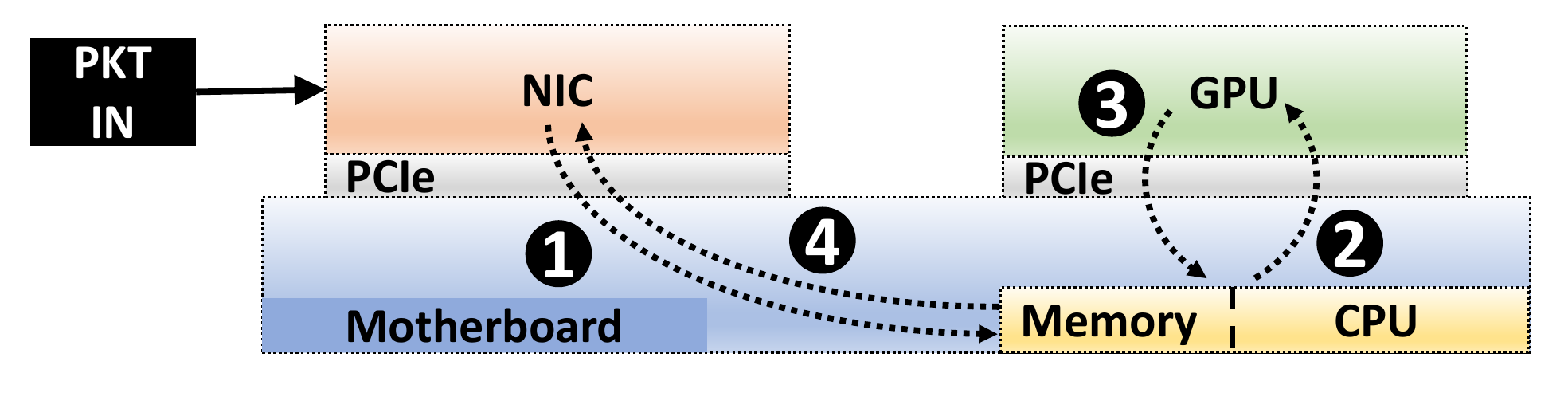}
    \vspace{-0.8cm}
    \caption{A system with a GPU working as NN accelerator. Enabling the NIC at performing NN inference could save up to 4 data transfers over the PCIe bus.}
    \label{fig:example-arch}
    \vspace{-0.2cm}
\end{figure}

In summary, the main contributions of this paper are:
\begin{itemize}
\item We present \tonic, an approach to accelerate NN inference for network traffic on commodity 
programmable NICs, and implement it on two different hardware targets, showing the flexibility of 
our approach and its applicability on off-the-shelf NIC hardware.
\item We share the three implementation designs: a microC-based implementation for the NFP4000 
network processor; a P4-based implementation for the P4-NetFPGA framework; and a Verilog hardware 
design for the NetFPGA;
\item We provide a compiler that generates fully functional P4 code from NN abstractions. The 
compiler targets both a software bmv2 switch and a P4 NIC, using the P4-NetFPGA framework.
\item We present three different use cases releated to network monitoring: (i) anomaly detection, 
(ii) traffic classification and (iii) network tomography and we study the trade-offs and benefits 
in offloading neural network inference on modern NICs.
\end{itemize}

\section{Motivation}
\label{sec:motivation} 


Given the wide availability of PCIe-attached off-the-shelf NN accelerators, i.e. GPUs, one may 
reasonably consider the use of such systems to offload a CPU from the processing of NN workloads, 
rather than a NIC. However: (i) NIC hardware is placed in a strategic position to reduce PCIe data transfer overheads; 
(ii) widely deployed NN architectures require only a small number arithmetic operations
per byte of loaded data, thus imposing little computational burden on the NIC; (iii)
commodity programmable NICs have a fair amount of computational power, which is potentially
available depending on the input traffic patterns. We detail these three observations in the rest
of the section and then present a motivating use case.

\subsection{Observations}
\noindent\textbf{I/O operations are expensive}.
Host-attached accelerators, i.e., GPUs, may introduce considerable data transfer overheads, 
as they are generally connected to a host system via the PCIe bus (Figure~\ref{fig:example-arch}). This rules them out as offloading engines for time-sensitive network applications.
For example, offloading the computation of NN inference to a GPU requires the input 
data to enter the host system (first PCIe transfer), then to reach the GPU (second PCIe transfer), and finally to transfer 
the result back to the CPU (third PCIe transfer). If the NN inference result is needed
for packet forwarding decisions, this may involve a potential additional transfer to the 
NIC (fourth PCIe transfer)\footnote{It is worth noting that although the newly designed 
GPUDirect RDMA (Remote Direct Memory Access) technology enables network devices to directly 
access GPU memory thus bypassing CPU host memory~\cite{gpudirect}, it cannot prevent data 
from crossing the PCIe.}.

\begin{figure}[t!]
    \centering
    \includegraphics[width=1\columnwidth]{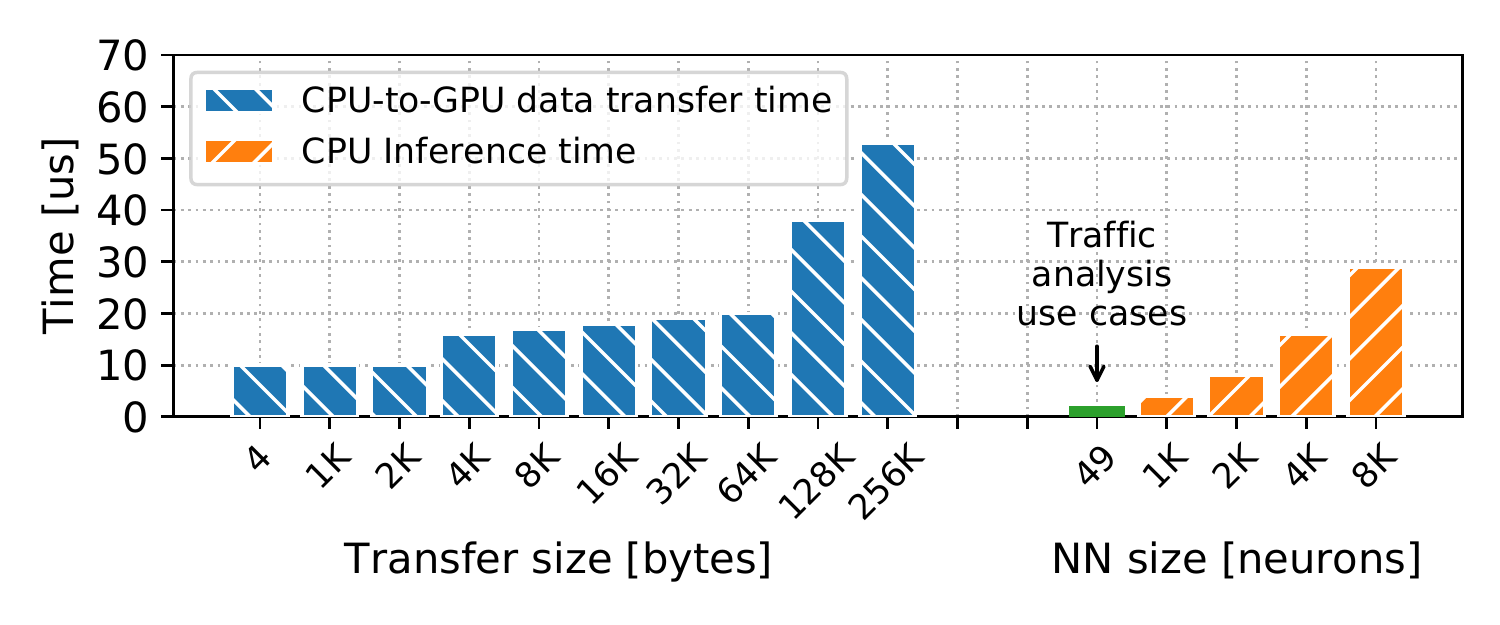}
    \vspace{-0.9cm}
    \caption{The PCIe RTT compared to the time needed by a single CPU core to run NN inference. A single PCIe RTT is larger than the NN inference time for small NNs.}
    \label{fig:io-cost}
    \vspace{-0.3cm}
\end{figure}

To better understand the overheads in place, we measured the time needed to transfer an
increasingly bigger input vector to a GPU through the PCIe and then retrieve back just 
the result from the inference.\footnote{We set this to 1B to minimize as much as possible the cost
of the I/O.} We compared this with the amount of time needed to directly run the inference 
on the CPU itself for a variable-sized NN (Figure~\ref{fig:io-cost}).\footnote{This test 
used an Intel Xeon E5-1630 v3 CPU (4 cores@3.7GHz) connected through a PCIe x16 v3.0 to 
an NVIDIA GTX 1080 TI 11GB.} The results show that transferring just few bytes of input 
vector and retrieving back the result from the GPU might already require 8-10\si{\micro\second}. 
To put this in perspective, executing relatively small NNs directly on the CPU takes as much 
as a single transfer. For instance, an efficient binary NN~\cite{BNN_NIPS2017} with about 2k 
neurons takes about 8\si{\micro\second} to be executed. Furthermore, depending on the specific 
use case, and as demonstrated in \S\ref{sec:use_cases} in the context of traffic analysis, 
the NN might need less than 50 neurons. In this scenario, the CPU only needs about 400\si{\nano\second} 
to run the inference, making inefficient the use of ML accelerators that need to 
be reached through PCIe. While batching the execution of multiple NNs together could be an 
effective way to reduce the data transfer overhead, it would also significantly impact the 
processing latency. For time sensitive workloads, NN inference is typically performed 
directly on the CPU, as it is a more effective solution than using a PCIe connected accelerators~\cite{hazelwood18}.
\begin{tcolorbox}
\textbf{Observation 1:} \textit{Running NN inference on a NIC could completely avoid 
	crossing the PCIe bus and help in the execution of time sensitive workloads.}
\end{tcolorbox}

\vspace{0.1in}
\noindent\textbf{NN processing is memory-bounded in many cases}. A NN is generally composed 
of several layers. Depending on the neurons' interconnections and operations, different types of 
layers can be defined, and, accordingly, different types of NN. \tonic implements \textit{fully connected}
layers (FCs), which are the building block of Multi-layer Perceptrons (MLPs). Other well known and widely used
NNs are Recurrent NNs (RNNs) and Convolutional NNs (CNNs). However, in 
classification and prediction tasks for formatted and tabular data, with low dimensions and lower 
number of target categories, MLPs are usually used, being fairly efficient. Indeed, despite the 
popularity of RNNs and CNNs, large industries such as Google and Facebook report that the large 
majority (e.g., above 60\%) of their NNs workloads requires the execution of MLPs~\cite{tpu, fbml}. 

MLPs, compared to other NNs, are a better fit to the constraints of networking devices, 
mainly due to the computation profile of FCs, which require much lower computation than other types of layers as demonstrated in the following test.
We used Intel's optimized Caffe~\cite{intelcaffe} to perform tests on a machine equipped with an 
Intel Xeon E5-1630 v3 CPU (4 cores@3.7GHz), and run VGG16~\cite{vgg} as test workload. VGG16 is a 
well-known CNN with a sequential structure, with convolutional (conv) layers followed by FCs, which 
allows us to clearly observe the different computation profiles of such layers. We use a single core 
to perform the computations, and pre-load the model in the local memory.\footnote{The core has 64KB of L1 
cache memory, 256KB of L2 cache and 10MB of L3 cache, which 
are essentially dedicated to the processing of such core, being the other 3 cores completely idle.} 
With \texttt{perf}, we monitor the CPU's performance counters for number of instructions, cycles 
and L3 cache misses. We use the Instructions per Cycle (IPC) rate as a proxy for the 
arithmetic intensity of the workload, i.e., the number of operations performed per byte 
of data loaded. Intuitively, high IPC and low L3 cache misses tell that the workload keeps most of 
the CPU's pipeline busy. Low IPCs and high L3 cache misses are a good indicator of memory-bound 
workloads, i.e., a small number of operations is executed for byte of loaded memory.

\begin{figure}[t!]
    \centering
    \includegraphics[width=1.0\columnwidth]{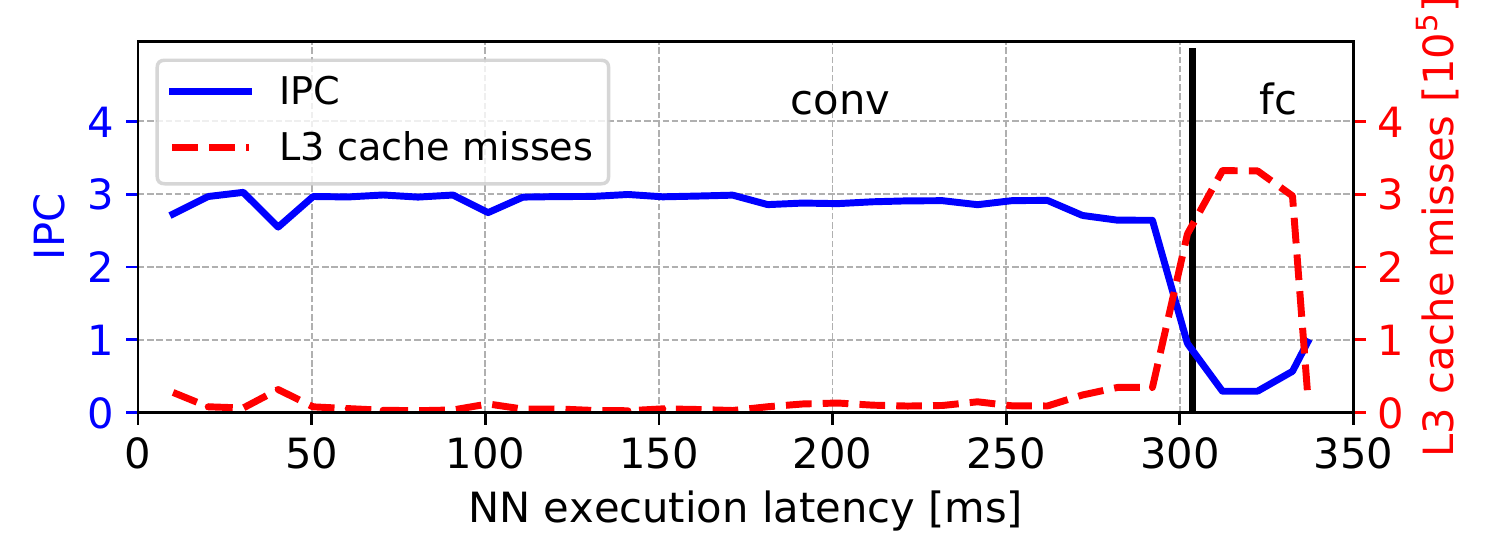}
    \vspace{-0.8cm}
    \caption{IPC and cache misses during the execution of the different layers of a VGG16 CNN. Unlike convolutional layers, fully-connected layers have a relatively low number of operations per byte of data loaded.} 
    \label{fig:normalized_per_layer_batched_inference}
    \vspace{-0.3cm}
\end{figure}

Figure~\ref{fig:normalized_per_layer_batched_inference} shows that the IPC is high when executing 
convolutional layers, while it is low during the execution of FC layers. The increase in L3 
cache misses shows that this happens due to the need to wait for data to be loaded from memory, 
confirming the relatively low arithmetic intensity of MLPs.
\begin{tcolorbox}
\textbf{Observation 2:} \textit{MLPs have relatively low arithmetic intensity. This makes hardware architectures optimized 
for fast memory accesses a good fit for their execution.}
\end{tcolorbox}

\begin{figure}[t!]
\begin{minipage}[t]{0.48\linewidth}
    \includegraphics[width=\linewidth]{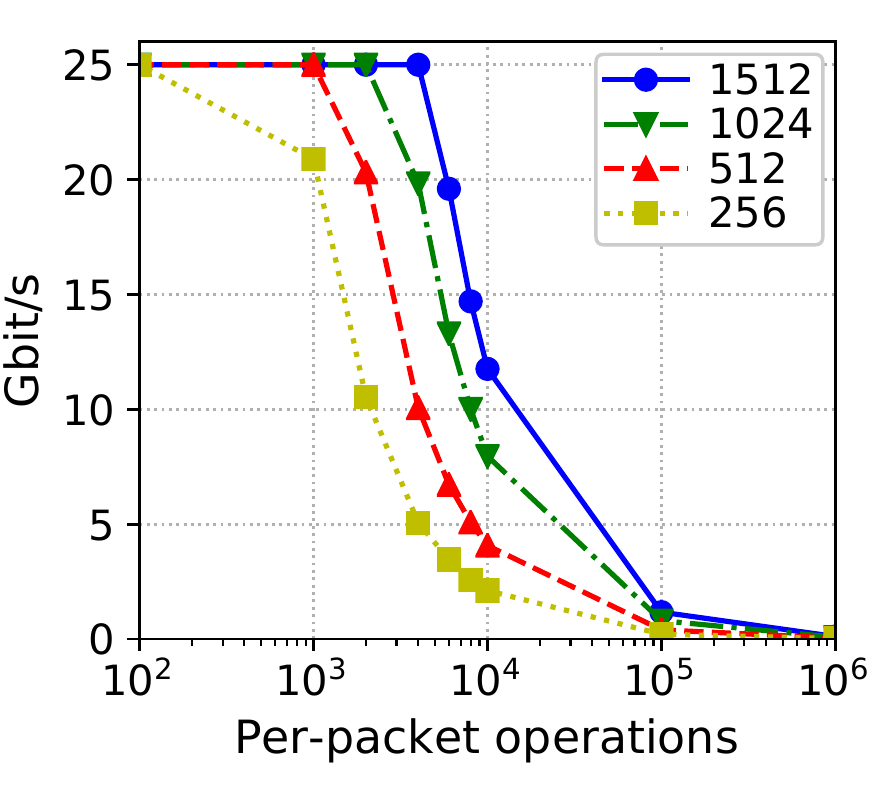}
    \vspace{-0.8cm}
    \caption{Forwarding throughput of a Netronome NFP when performing an increasing number of arithmetic operations per-packet. With larger packet sizes many operations can be performed, without affecting the forwarding throughput.}
    \label{fig:tput_operations}
    \vspace{-0.3cm}
\end{minipage}%
    \hfill%
\begin{minipage}[t]{0.48\linewidth}
    \includegraphics[width=\linewidth]{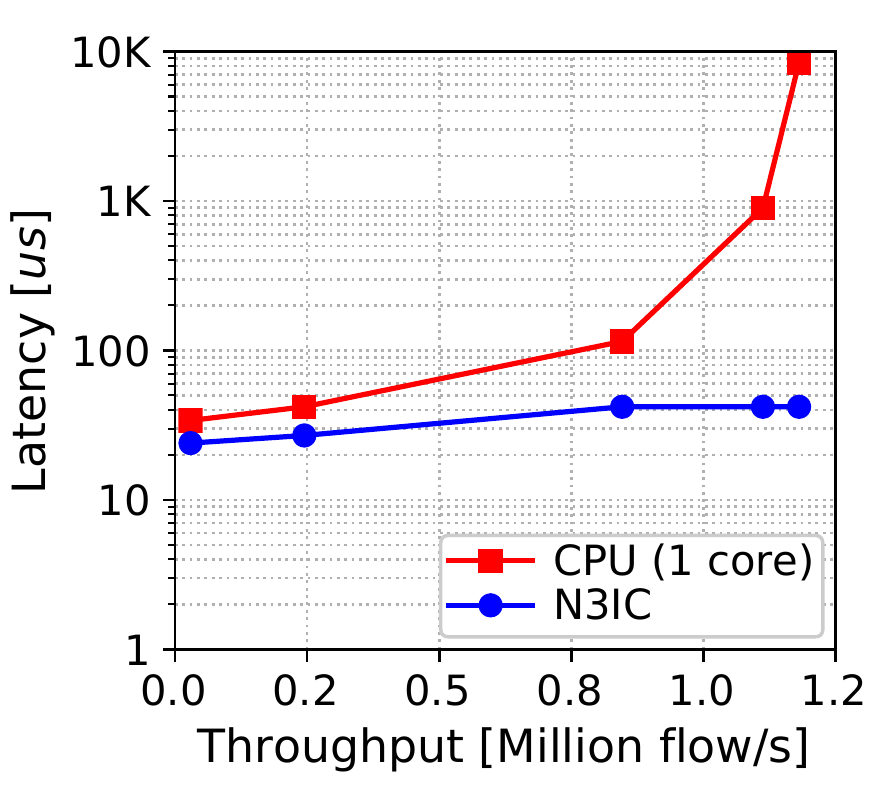}
    \vspace{-0.8cm}
    \caption{Comparison of the processing latency for A CPU-based executor and \tonic, when an increasing number of network flows/s has to be analyzed. By running on the NIC, for higher throughput values \tonic can provide a 100x lower latency.}
    \label{fig:motivation}
    \vspace{-0.3cm}
\end{minipage}
\end{figure}

\vspace{0.1in}
\noindent\textbf{NICs processing load depends on traffic patterns}. A NIC handling a 
bandwidth of 40Gb/s needs to process about 3 million packets per second (Mpps) when packets are 1500B
in size. In contrast, for the same bandwidth, a NIC would need to process 15Mpps of size 256B, 
which is a 5x bigger processing load. Given that the hardware of NICs is often designed 
to handle worst case traffic scenarios, i.e., processing many small network packets, this creates 
opportunities to run additional operations without paying operational penalties. To illustrate this, 
we show in Figure~\ref{fig:tput_operations} the throughput of a Netronome NIC when performing 
an increasingly higher number of per-packet operations, i.e, integer arithmetic over the received 
packet, beside the common parsing and forwarding tasks, when loaded with 25Gb/s of constant bit-rate 
traffic with packet sizes of 512B, 1024B and 1500B. The figure shows that as the packet size increases, 
the NIC can perform more per-packet operations in addition to its regular forwarding tasks, without 
affecting the forwarding throughput. Specifically, considering an average case of 512B input packets, 
typical for a data center scenario~\cite{roy15}, the available budget is of 10K operations per-packet 
before trading with operational performance. Notably, the processing budget grows by ten times when 
the packets size doubles.
\begin{tcolorbox}
\textbf{Observation 3:} \textit{The relatively small arithmetic intensity of MLPs aligns well with the potentially available processing power of a NIC.}
\end{tcolorbox}

\subsection{A motivating use case}
NNs are a relevant tool for traffic classification~\cite{moore05,liang19} and anomaly detection~\cite{kathareios17}.
Usually, the end-to-end process requires the reception of packets, then the calculation of traffic 
statistics that will ultimately feed an NN in charge of producing the final result.
In this scenario, a NIC usually offloads the 
CPU only for the network flows statistic collection. The NN inference 
has to be entirely run on the host's CPU, which therefore has to continuously fetch the traffic statistics from 
the NIC's hardware to perform the NN inference. Although the CPU no longer has to run expensive operations such as packet reception and statistics collection, the 
NN inference can already occupy an entire CPU core, even when performing optimized batched computation of the per-flow statistics.

In Figure~\ref{fig:motivation}, we show the performance achieved by using a single CPU core 
for NN inference. Here, the traffic analysis can scale to process up to 1.2M network 
flows per second, but only when performing input batching. Unfortunately, batching increases 
the processing latency from 10s of \si{\micro\second}, for a batch of size 1, to 10s of ms, 
for a batch of size 10k. This hinders the implementation of time-sensitive applications relying 
on quick results. \tonic, instead, without using expensive hardware, allows the system to 
completely offload most of the computation using available NIC's resources, thereby avoiding 
data movements, freeing CPU cores, providing timely results and reducing processing latency by up to 100x.

\section{Neural Networks on the NIC}
\label{sec:solution}
In this section we describe the challenges of supporting NN inference in the 
data plane of a commodity programmable NIC, and how we overcame them with \tonic.

\vspace{0.1in}
\noindent\textbf{\underline{Challenge 1:} The need for small NN models:} 
A typical NIC has at most few 10s of MBs of fast on-chip SRAM memory~\cite{netronome,tilencore,bluefield}, 
which is commonly used to store the data required for packet processing. 
Small NNs, which are common in networking use cases (\S\ref{sec:use_cases}), 
could fit their model's parameters in such space. However, the on-chip 
memory needs to also host forwarding and policy tables. Such tables can 
easily take several MBs of memory, leaving little space available for the 
NNs. An important observation is that NNs, being inherently a tool of 
approximate computing~\cite{Hornik}, can effectively take advantage of 
compression and quantization techniques. These techniques reduce the number 
of bits used to represent the NN's parameters, e.g., 8b instead of 32b, and 
are already widely applied to improve the efficiency of NN accelerators. 
Recent approaches have pushed quantization even further, using just few 
bits per parameter~\cite{Lee_2017, XNOR-net, ZhouNZWWZ16}. Effectively, 
this provides opportunities to reduce memory requirements by 2-8x even when compared to already efficient 8b representations. Although 
there is a trade-off between memory requirements and NN accuracy, for many use cases
the drop in accuracy is usually tolerable~\cite{Lee_2017}.

\vspace{0.1in}
\noindent\textbf{\underline{Challenge 2:} The need for a low complexity algebra:} 
A second major constraint is related to the arithmetic capabilities of networking 
devices. Networking hardware often only has the ability to perform 
bitwise logic, shifts and simple integer arithmetic operations~\cite{sivaraman16}. 
Implementing the operations required by a NN, such as multiplications, may not always be possible. Certain quantization techniques help in addressing 
this issue, since they can simplify the arithmetic operations required to compute 
a NN. Indeed, the already mentioned 8b quantization is used by the TPU~\cite{tpu} 
to enable the use of integer multiplications in place of floating point. More 
interestingly, some quantization techniques can even completely change the nature 
of the operations required to compute a NN. For example, log domain quantization~\cite{Lee_2017} 
uses log base-2 values to represent NN parameters, which enables replacing 
multiplications with simple shift operations.

\vspace{0.1in}
\noindent\textbf{\underline{Challenge 3:} The need for hardware compatibility:} 
The number of currently available quantization techniques is fairly large, e.g,~\cite{bnn,Lee_2017,XNOR-net,ZhouNZWWZ16,Han2015DeepCC,CaiHSV17,Zhu2016TrainedTQ,Li2016TernaryWN}. 
However, our solution space is reduced by a third final constraint: networking 
devices have quite heterogeneous hardware architectures, ranging from SoCs, FPGAs, 
embedded CPUs to switching ASICs. Thus, the selected quantization technique should 
guarantee a wide hardware compatibility. Effectively, this translates into selecting 
a quantization approach that uses arithmetic operations widely supported in the 
above architectures.

\begin{algorithm}[!h]
        \SetKwInOut{Input}{Input}
        \SetKwInOut{Output}{Output}
        \SetKw{KwBy}{by}
        \Input{$x$ input vector, $w$ weights matrix, $n$ num. of output neurons;}
        \Output{$y$ output vector}

        $block\_size \gets 32 $\;
        $\textbf{assert}( n \mathrel{\%} block\_size \mathrel{{=}{=}} 0 ) $\;
        $sign\_thr = (\textbf{len}(x)\mathrel{{*}}block\_size)/2 $\;
        $\textit{y}[n/block\_size] \gets \{0\} $\;

        \For {$neur\gets0$ \KwTo $n$ \KwBy $1$}{
                $tmp\gets0$\;
                \For {$i\gets0$ \KwTo $\textbf{len}(x)$ \KwBy $1$}{
                        $tmp \mathrel{{+}{=}} \textbf{popcnt}(w[neur][i] \odot x[i])$\;
                }
                \If {$tmp \mathrel{{>}{=}} sign\_thr $}{
                        $tmp\_out \mathrel{{|}{=}} (1 << (neur \mathrel{\%}block\_size ))$\;
                }
                \If {$(neur+1) \mathrel{\%} block\_size \mathrel{{=}{=}}0 $}{
                        $y[neur]\gets tmp\_out$\;
                        $tmp\_out\gets 0$\;
                }
        }
        \caption{FC processing function. Weights and inputs are in groups of \texttt{block\_size}.}
        \label{alg:bnn_processing}
\end{algorithm}

\vspace{-0.2in}
\subsection{The solution adopted by \tonic}
With the previous challenges in mind, we finally select \textit{binary neural networks} 
(BNN) as the \tonic's quantization technique to implement NN in the network data 
plane of a NIC. A BNN uses only 1b to represent inputs and weights~\cite{XNOR-net,trainingBNN}, 
and replaces multiplications with a combination of simpler operations, i.e., 
bitwise XNOR, and population count (\texttt{popcnt}). Compared to other solutions, 
such as log domain quantization, which requires shift operations, BNNs provide better 
opportunities for parallelizing the computation. Finally, despite the significantly lower 
computation and memory complexity, BNNs can still achieve remarkable results. 
In many tasks, BNNs provide prediction accuracy only 3-10\% points lower than 
the accuracy of the much heavier non-quantized models~\cite{BNN_NIPS2017}. While 
this may be a significant drop for some use cases, it still enables such models 
to be used in a large set of applications, as we will discuss more thoroughly 
in \S\ref{sec:use_cases}. 
Precisely, starting from the original NN, we first applied the binarization technique from 
Courbariaux and Bengio~\cite{trainingBNN}, which is based on a canonical back-propagation 
algorithm. This solution ensures that the NN's weights converge to values included in the 
[-1, 1] range, and that are normally distributed around the 0. This helps in losing 
only little information when the real weight values are mapped to just two values, i.e., 
0 and 1~\cite{trainingBNN}. In fact, the MLP's weights obtained after training are still 
real numbers, and an additional final step is required to obtain the binarized weights. 
This is done by setting the NN weigths to 0 if their value is negative or 1 otherwise.

We implemented the processing of a BNN with Algorithm~\ref{alg:bnn_processing}, which 
parametrizes the BNN operations according to the variable \texttt{block\_size}. This 
corresponds to the largest contiguous set of bits on which an operation can be performed 
in the target hardware executor. For instance, a modern CPU could have a \texttt{block\_size} 
of 64, corresponding to the CPU's registers size. Specifically, for each of the NN's 
neurons, the algorithm performs (1) a XNOR of the input with the weights vector, (2) a 
population count operation and (3) a comparison. The comparison checks if the result of 
the population count is smaller than half the number of neuron's inputs, in which case 
the neuron's output is 1. 

\begin{figure}[t!]
    \centering
    \includegraphics[width=1\columnwidth]{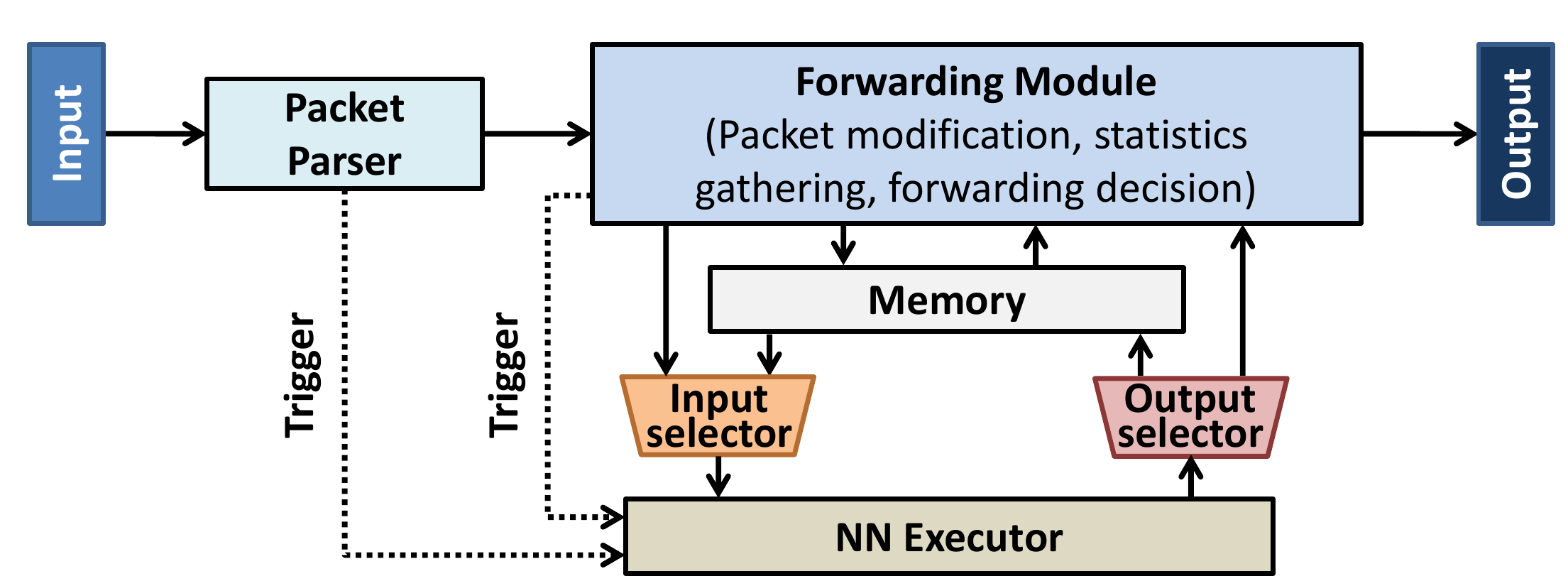}
    \vspace{-0.8cm}
    \caption{\tonic logical architecture.}
    \label{fig:architecture}
    \vspace{-0.3cm}
\end{figure}

\subsection{\tonic architecture}
The above described BNN algorithm is integrated in the NIC data plane as depicted  
in Figure~\ref{fig:architecture}. The NN Executor is separated from
the forwarding module, although some implementations may allow both instances to be 
integrated. The NN Executor is triggered either by the reception of a network packet, 
or directly by the forwarding module in charge of the regular packet 
processing. In fact, the forwarding module may collect network flows statistics and 
only trigger the NN Executor when enough packets have been received.

The input selector allows to choose the input of the NN executor between a 
packet field or a specific memory area, e.g., containing collected flow 
statistics. The output selector is used instead to write the inference output 
either to a packet field, or to a specified memory location. 
When the input and output selectors are configured to read or to write to 
a packet field, the NN Executor works as an inline module.

\section{Implementations}
\label{sec:impl}
In this section we describe the implementation of \tonic on two different programmable 
hardware targets: a SoC-based (NFP4000 Netronome~\cite{netronome}) and an FPGA-based 
(NetFPGA~\cite{netsume}) NICs. We first show the implementation of \tonic using only the 
currently available NICs' primitives accessible through high level programming, i.e., 
microC and P4. We then describe the design of a new hardware module that implements a 
dedicated BNN executor using Hardware Description Language (HDL). This 
provides a pathway to enable the use of our solution as a primitive for high-level languages.

\subsection{SoC NIC: Netronome NFP4000}
\begin{figure}[t!]
    \includegraphics[width=\columnwidth]{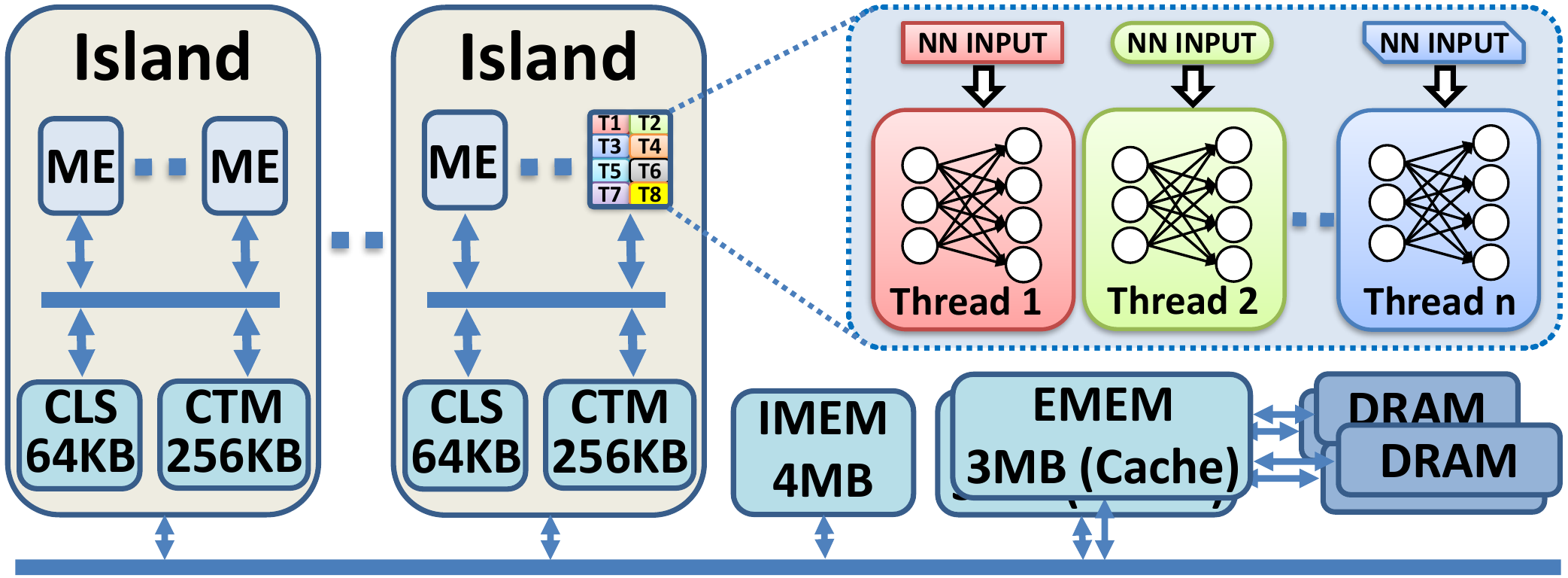}
    \vspace{-0.8cm}
    \caption{The architecture of a Netronome NFP4000's programmable blocks and the NN processing with \tonic-NFP.}
    \label{fig:netronome}
    \vspace{-0.3cm}
\end{figure}

The NFP4000 architecture comprises several different hardware blocks. Some of them are 
dedicated to network-specific operations, e.g., load balancing, encryption. Others are 
binded to programmable components that are used to implement custom packet operations. 
The architecture is shown in Figure~\ref{fig:netronome}, and comprises tens of independent 
processing cores, which in Netronome terminology are named micro-engines (MEs). MEs are 
programmed with a high level language named \textit{micro-C}, a C dialect. Each ME has 
8 threads, which allow the system to efficiently hide memory access times, e.g., context 
switching between threads as they process different packets. MEs are further organized in 
islands, and each island has two shared SRAM memory areas of 64KB and 256KB, called CLS 
and CTM, respectively. Generally, these memory areas are used to host data required for 
the processing of each network packet. Finally, the chip provides a memory area shared 
by all islands, the IMEM, of 4MB SRAM, and a memory subsystem that combines two 3MB SRAMs, 
used as cache, with larger DRAMs, called EMEMs. These larger memories generally host 
forwarding tables, access control lists and flow counters. MEs can communicate and 
synchronize with any other ME, irrespective of the location. Communications across islands 
take longer and may impact the performance of a program.

When developing \tonic on this hardware platform we have to share the MEs and memory 
resources between tasks, thus, we had to strike the right balance between the needs of 
quickly forwarding network packets and running NN inference. For both processing tasks 
the main bottleneck is the memory access time. Thus, selecting the memory area to store 
NN's weights has played a major role in our design. 

If the NN is small, as for the use cases considered in this paper (\S\ref{sec:use_cases}), 
it is worth considering the fastest available on chip memories, i.e., the CTM and CTS, 
with an access time of less than 100ns~\cite{netronome}. However, the CTM memory is usually 
dedicated to packet processing tasks, being the memory used by the NFP to store incoming 
packets and making them available to the MEs. Thus, using the CTM may impact packet processing 
and should be avoided. 
Because of this, we decided to load the NN's weights at configuration time in the CLS memory.
At boot time, each of the MEs' threads registers itself to be 
notified of packets reception, with the NFP taking care of distributing packets to threads 
on a per-flow basis. This is a standard approach when programming the NFP. Thus, whenever 
a new packet is received, the NFP copies its content in an island's CTM, and notifies one 
of the island's threads to start packet processing. The notified thread performs regular 
packet processing tasks, such as parsing, counters update, forwarding table lookups. If 
a trigger condition is verified, the thread starts the processing of the configured NN. 
Typical conditions could be the arrival of a new flow, the reception of a predefined number 
of packets for a given flow, the parsing of a given value in a packet header. 
To run the NN, a thread performs Algorithm~\ref{alg:bnn_processing}, with input and weights
packed in 32b integers, i.e., \texttt{block\_size} is 32. As a consequence, multiple threads 
can perform NN inference in parallel (Figure~\ref{fig:netronome}).

When NNs are larger, running them in a single thread would take long, because the NN has 
to be stored in slower off-chip memory. In this case, using multiple threads to run 
a single inference is more effective, even if synchronization among them incurs in some overhead. 
We leave the discussion of this trade-off to the Appendix.

\subsection{From Neural Networks to P4 and to NetFPGA}
P4~\cite{p4} is a domain-specific, platform agnostic language for the programming 
of packet processing functions. We designed a compiler that transforms an NN 
description into a \tonic implementation described with P4. 
In principle, the P4-based implementation allows us to separate the \tonic solutions 
from the underlying hardware-specific details, thus generalizing our approach.
However, as we will discuss at the end of the section, the target hardware architecture 
has still an important impact on the final implementation.

\vspace{0.1in}
\noindent\textbf{Compiling NN to P4.} Our compiler, \nnp, takes as input a 
NN description, i.e., number of layers with corresponding number of neurons, 
and generates $P4_{16}$ code for a generic P4 target based on the PISA architecture. 
PISA is a spacial forwarding pipeline architecture, with a number of match-action 
units (MAUs) in series. A packet header vector (PHV), containing both the input 
packet and metadata information, is passed through the MAUs to perform the 
programmed processing tasks. Each MAU combines a table memory structure, for 
quick lookups using the PHV fields, with arrays of ALUs that perform operations 
on such fields. The code generated by \nnp implements a function, on top of 
the PISA architecture, which reads the input value from the PHV, performs the 
NN execution and writes back to a PHV's field the result of the computation. 
The NN weights are stored in the MAUs' fast memories to enable runtime reconfiguration.
The generated P4 code also includes headers definition, parser, de-parser and 
control blocks. The code can therefore be easily extended to integrate with 
any other required packet processing function.

\begin{algorithm}[!h]
    \SetKwInOut{Input}{Input}
    \SetKwInOut{Output}{Output}
    \SetKw{KwBy}{by}
    \Input{$n$ input number;}
    \Output{$c$ output counter}
    $L \gets log_2B $\;
    $bits[L] \gets \{1, 2, 4, ..., B/2\} $\;
    $masks[L] \gets \{\texttt{01}\textbar_{B/2}, \texttt{0011}\textbar_{B/4}, \texttt{00001111}\textbar_{B/8}, ..., \texttt{0}\textbar_{B/2}\textbar\textbar\texttt{1}\textbar_{B/2}\} $\;
    $c \gets n $\;
    \For {$i\gets0$ \KwTo $L-1$ \KwBy $1$}{
            $c \gets (c \mathrel{\&} masks[i]) + ((c >> bits[i]) \mathrel{\&} masks[i])$\;
    }
    \caption{popcount implementation. B is the number of bits required to represent $n$. $\texttt{X}\textbar_{y}$ indicates the $y$-times concatenation of the binary number \texttt{X} and $\texttt{Z}\textbar\textbar\texttt{W}$ is the concatenation of the binary numbers \texttt{Z} and \texttt{W}.}
    \label{alg:popcnt}
\end{algorithm}

The basic operations needed to implement Algorithm~\ref{alg:bnn_processing} are 
(1) XNOR, (2) popcount and (3) SIGN function.
Executing a XNOR and a comparison (SIGN) is readily supported by the P4 language. 
Unfortunately, the popcount operation is not. The main issue is that its execution 
time depends on the input size, which makes popcount difficult to implement 
in networking hardware, and therefore not supported in the PISA architecture. 
To overcome this issue using only current P4 primitives, 
we adopted the solution proposed in~\cite{beeler1972hakmem} and reported in Algorithm~\ref{alg:popcnt}. 
The idea is to implement the popcount by combining basic integer arithmetic and logic 
operations in a tree structure whose depth is dependent on the input size. A tree 
structure is easily parallelizable and allows the distribution of 
computations in the pipeline, with the processing of different tree's levels assigned 
to different pipeline's stages.

\begin{figure}[t!]
        \centering
        \includegraphics[width=1\columnwidth]{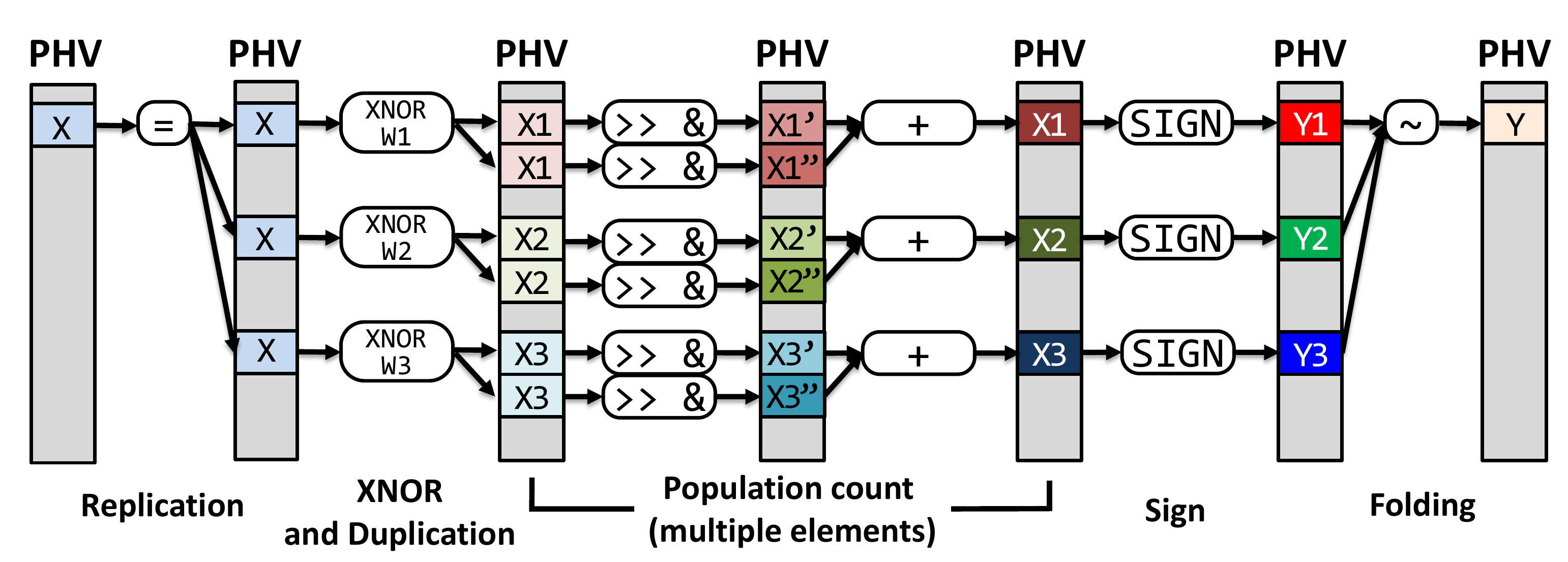}
	\vspace{-0.8cm}
        \caption{The logical steps required to implement a BNN using a PISA architecture.}
        \label{fig:p4algo}
        \vspace{-0.3cm}
\end{figure}

Overall, the processing includes five steps, each one mapped to a logical pipeline stage, 
except for the popcount which requires multiple stages, depending on the input size (cf. Figure~\ref{fig:p4algo}).
In the first step, the NN input is replicated in as many PHV fields as the number of neurons 
to exploit the parallel processing on multiple packet header fields. Specifically, this 
corresponds to an unrolling (or partial unrolling) of the first \texttt{for} cycle of 
Algorithm~\ref{alg:bnn_processing}. In the second step, each field, containing a copy of 
the NN input, is XNORed with the corresponding weight. The resulting value 
is further duplicated to additional fields to implement the shift, 
AND and sum as described in Algorithm~\ref{alg:popcnt}. The outcome of each popcount is then 
compared with a threshold to implement the SIGN function, whose result is the output of 
each neuron. Finally, the resulting bits, stored in one PHV field for each neuron, are 
folded together in a single field. Depending on the NN depth, 
\nnp  replicates and concatenates the described operations as many times as the number of 
layers to obtain the complete MLP execution.
We functionally tested the generated P4 implementations using the software target bmv2.

For hardware targets, it is worth noticing that the PHV size limits the 
number of neurons the pipeline can execute in parallel. 
This is due to the need to replicate the input in the PHV to 
enable parallelism at the MAU level. Furthermore, the number of neurons executed in 
parallel by a single MAU is also limited by the maximum amount of memory a MAU can read 
in a single operation.

\vspace{0.1in}
\noindent\textbf{Porting P4 code to NetFPGA.}
The NetFPGA is a programmable hardware platform with 4x10GbE Ethernet interfaces 
incorporating a Xilinx Virtex-7 FPGA. We implemented \tonic on top of the reference 
NIC project provided with the main NetFPGA-SUME code base.\footnote{\url{https://github.com/NetFPGA/NetFPGA-SUME-public/wiki/NetFPGA-SUME-Reference-NIC}}
We used the P4-NetFPGA workflow~\cite{ibanez19} to port the generated target-independent
P4 code to the NetFPGA platform. The P4-NetFPGA workflow is built upon the Xilinx 
P4-SDNet~\cite{xilinx19} compiler and the NetFPGA-SUME code base. It translates P4 code 
to HDL (Verilog), and integrates it within the NetFPGA pipeline. 

The P4-NetFPGA workflow required several adaptations to \nnp, in order to meet the 
FPGA resources and timing constraints. First, the P4-SDNet compiler does not support 
\texttt{if statements} among the operations of a MAU. Thus, we replaced all the if 
statements required by the SIGN function using a combination of bitwise logic operations 
and masks. Second, MAUs use the CAM IP core from Xilinx to implement lookup tables, 
which restricts the maximum width size that can be used for each entry. Consequently, 
a maximum of 32B can be fetched from memory every time a table is called, limiting the 
number of neuron weights that could be loaded in parallel by each table. To overcome 
this issue we had to write the weights as constant values in the MAU's operations code, 
effectively trading the possibility to perform runtime reconfiguration with the ability 
to compute more neurons in parallel. Finally, P4-SDNet is capable of performing a large 
number of operations on a field in a single MAU. This is in contrast with ASIC targets, 
which are instead usually constrained to execute a single operation per MAU~\cite{sivaraman16}. 
This allowed us to describe several steps of a BNN computation in a single MAU, thus 
reducing the number of MAUs required to implement the BNN computation.

\subsection{BNN inference primitive}

\begin{figure}[t!]
    \centering
    \includegraphics[width=1\columnwidth]{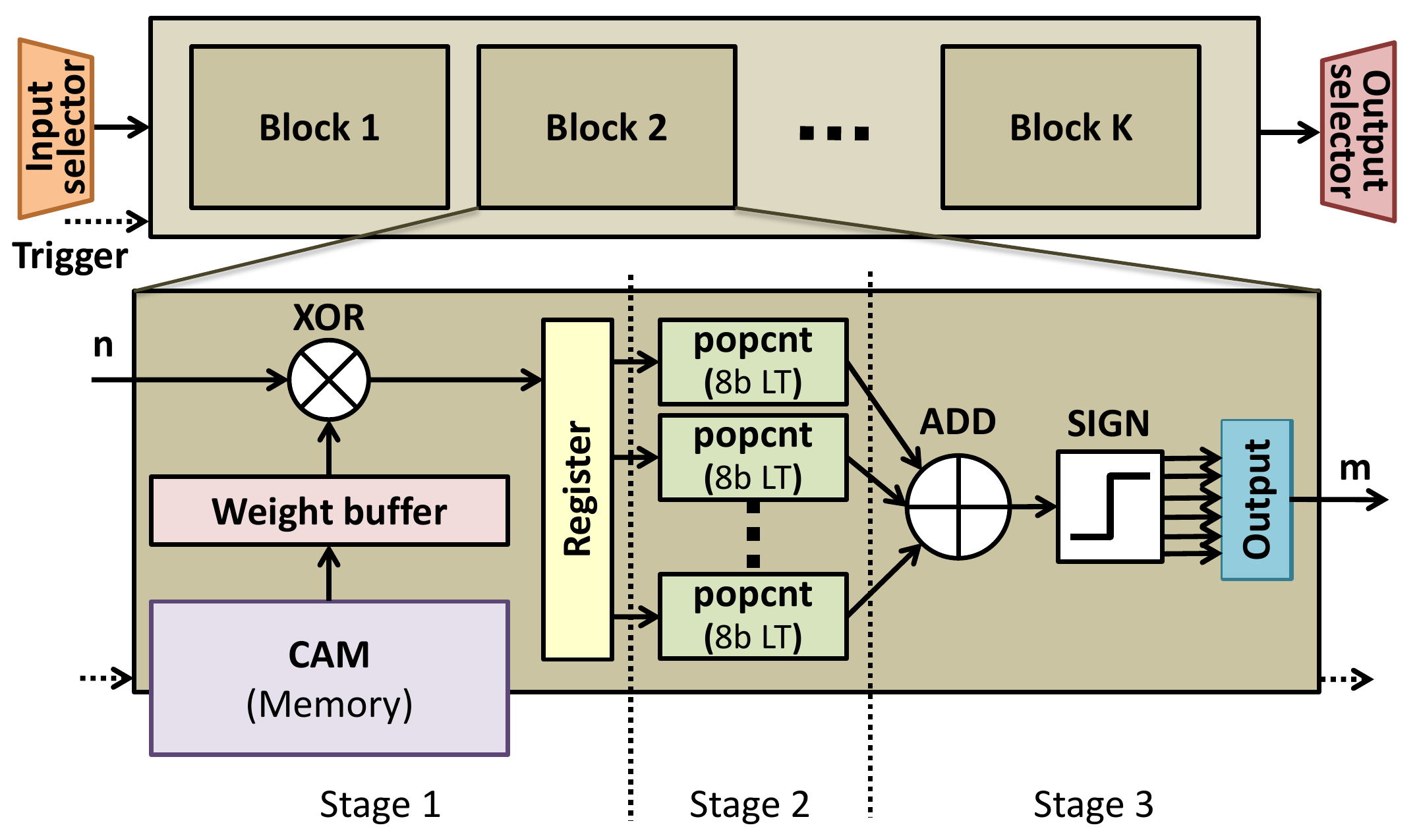}
    \vspace{-0.8cm}
    \caption{Hardware design of the NN Executor module.}
    \label{fig:fpga_design}
    \vspace{-0.3cm}
\end{figure}
               
Finally, we designed a dedicated NN executor for the NetFPGA in HDL, which allows the 
NIC to expose BNN inference as an offloading primitive. Figure~\ref{fig:fpga_design} 
shows the architecture of our NN executor as per Algorithm~\ref{alg:bnn_processing}.
The module is composed by multiple blocks. Each of them performs the computation of a 
single NN layer, and can be parametrized providing the sizes $n$ and $m$ for the input 
and output vectors, respectively. Together, the blocks build an NN Executor for specific 
NN architectures. For instance, three of these blocks are required to build a 3 
layers MLP. The NN layer weights are stored in the FPGA on-chip memories, i.e., Block 
RAM (BRAM). The BRAMs are organized as tables with a number of rows strictly 
dependent to the amount of neurons and with a width of 256b. Each row can be read in 
2 clock cycles and, depending on the size $n$ of the input vector, can store one or 
multiple weights, e.g., 1x256b or 16x23b. The BRAMs are shared by all the blocks of 
a NN Executor module.

A single block is a pipeline of three stages. The first reads the weights from the BRAM 
and performs the XNOR with the input. The second performs a first step of the 
popcount. Here, we create Lookup-Tables (LTs) of 256 entries each, in order to associate 
one 8b integer (address) to the corresponding population count value. Each block has 
$n/8$ of these LTs. As a consequence, for a 256b input we create 32 LTs that operate in 
parallel. In the last stage, the LTs outputs are summed together, the sign function is 
applied on the final sum and its result is stored in one of the $m$ bits of the output 
register. If multiple weights are placed in a single BRAM's row, the module performs the 
execution of several neurons in parallel.

\section{Use cases}
\label{sec:use_cases}

\begin{figure}[t!]
\begin{minipage}[t]{0.48\linewidth}
         \includegraphics[width=\linewidth]{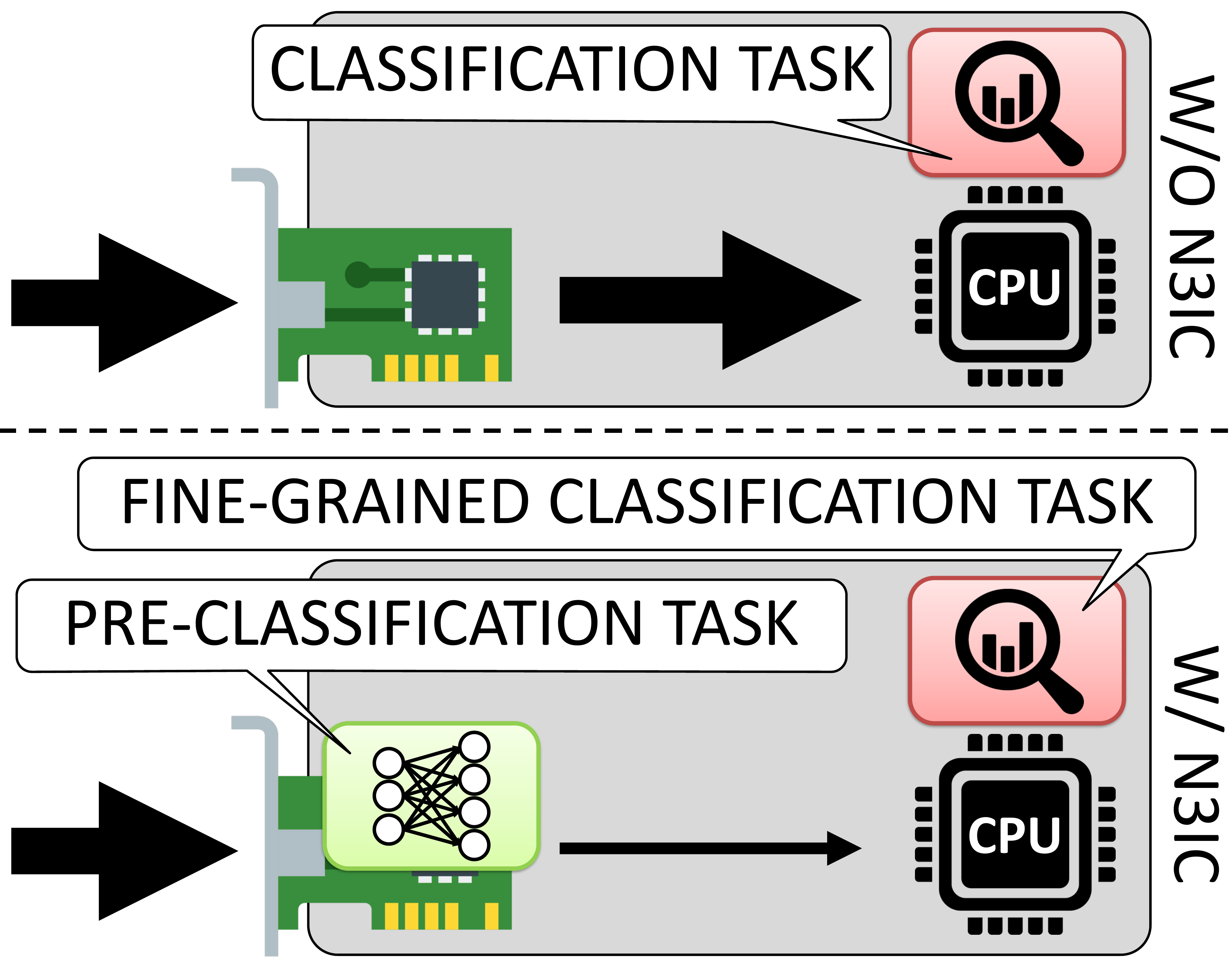}
    \vspace{-0.8cm}
         \caption{For traffic analysis use cases, \tonic can split the classification task and work as a flow shunting system, thereby increasing application's scalability.}
         \label{fig:usecase_tc}
    \vspace{-0.3cm}
\end{minipage}%
    \hfill%
\begin{minipage}[t]{0.48\linewidth}
         \includegraphics[width=\linewidth]{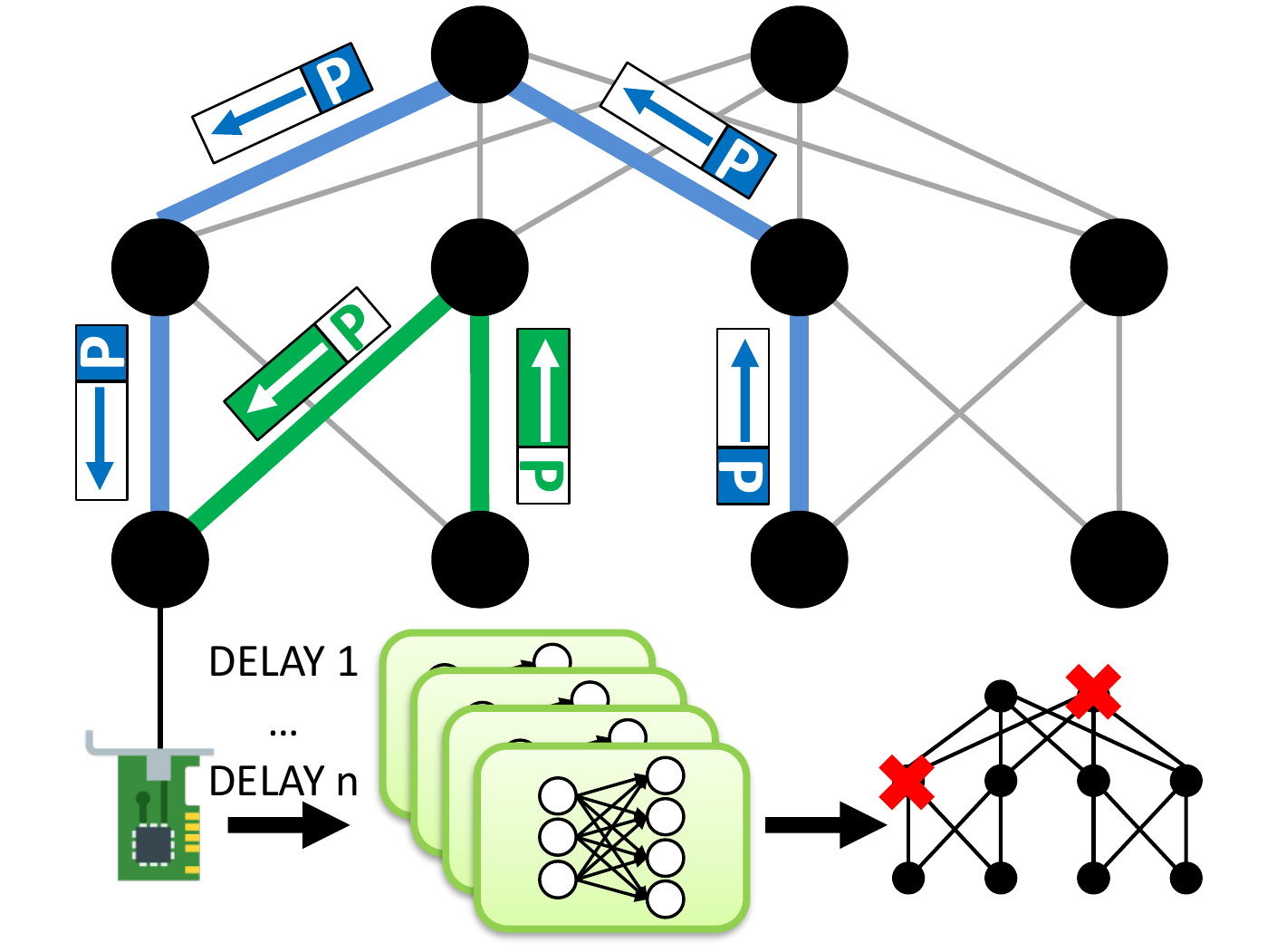}
    \vspace{-0.8cm}
         \caption{\tonic enables the real-time implementation of SIMON~\cite{geng19}, using NNs in the NIC to identify congested queue from probes' one-way delays.}
         \label{fig:usecase_simon}
    \vspace{-0.3cm}
\end{minipage}
\end{figure}

In this section, we show how \tonic can benefit applications from both backbone and datacenter networks. 
Table~\ref{tab:use-cases} summarizes the three implemented use cases. More details are provided in the 
Appendix.

\vspace{0.1in}
\noindent\textbf{\#1: Traffic classification.}
Backbone network operators use traffic classifiers to make a best-guess about the type of traffic 
carried by a flow, in order to assign priority classes and improve user experience~\cite{karagiannis05}. 
This operation is often carried at the network edge~\cite{tyson17} using software middleboxes for 
enhanced flexibility~\cite{martins13,martins14,sherry15} at the cost of reduced performance~\cite{sekar12,barbette15,katsikas18}. 
In this context, \tonic can be used to reduce the amount of traffic processed by the software 
classifier by performing a pre-classification task to reduce the pressure on the host CPU (See Figure~\ref{fig:usecase_tc}).
 
To illustrate this use case, we used the UPC-AAU dataset~\cite{bujlow2015independent}. The dataset 
has more than 750K flows, including 76 commonly used applications and services, which account for 
55GB of packet-level data. Here, \tonic can be used to classify the traffic class that contributes 
the most. In the UPC-AAU dataset, this corresponds to the identification of P2P traffic. With this
approach, \tonic would be in charge of directly classifying P2P traffic, while passing the rest 
to the software middlebox running on the host system for more fine-grained classification. That is, 
\tonic can be used as an effective flow-shunting system. We trained a binarized MLP to classify 
traffic in two classes (i.e., P2P and other), and deploy it with \tonic.  On this classification 
task, a binarized MLP with 32, 16, 2 neurons is capable of achieving 88.6\% accuracy, while requiring 
only 1KB of memory.

\vspace{0.1in}
\noindent\textbf{\#2: Anomaly detection.} Anomaly detection is the practice of analyzing 
traffic to seek anomalies and reveal suspicious behavior~\cite{zeek}. Unfortunately, to cope with the ever-increasing large traffic 
volumes, operators only perform analysis on randomly chosen samples of the flows~\cite{tilmans18}. 
With \tonic, instead, we can allow software middleboxes deployed at the edge network
to perform continuous flow-level analysis 
for all the flows, without impacting the resource usage of the host CPU.

For this use case we used the the UNSW-NB15 dataset~\cite{UNSW_NB15_paper}, which provides over 170K
network flows labeled in two main categories, i.e., good, bad, and their flow-level
statistics. A regular MLP with 3 FCs of 32, 16, 1 neurons achieves 90.3\% accuracy on this dataset, 
requiring 35KB of memory.
Using a similar binarized MLP (32, 16, 2 neurons and 1KB in total of memory) with \tonic achieves an 85.3\% 
accuracy, instead. While the \tonic NN gives a 5\% lower accuracy, it is applicable to large volumes of 
traffic, enabling a type of security monitoring that would not be economically viable otherwise.

\vspace{0.1in}
\noindent\textbf{\#3: Network Tomography.}
In datacenter networks, end-host based network monitoring approaches are being explored for their ease of 
deployment and lightweight operations~\cite{moshref16,ghasemi17,geng19}. For example, SIMON~\cite{geng19} 
periodically sends probe packets to measure datacenter's network paths delays. This allows to accurately 
infer congestion points and the size of the related queues. SIMON uses MLPs to speed-up the inference tasks. 
However, it relies on GPUs to run the MLPs, which makes the tool better applicable for debugging than for 
creating a feedback loop between measurement and control, i.e., for path selection. Here, notice that according 
to~\cite{geng19} the probe periodicity depends on the fastest link speed. For instance, in the presence of 
10Gb/s links, probes have to be sent every 1ms, while for 40Gb/s links this lowers to 0.25ms. As a consequence, 
for this use case to work at higher link speeds and in real time, the execution latency has to be lower than 
the probe sending periodicity. This can be challenging considering I/O overheads and MLP processing time
as previously discussed in \S\ref{sec:motivation}. With \tonic, we implemented a modified version of SIMON 
that can quickly identify congestion points, and timely notify them to the network control plane. 

\begin{table}[]{\footnotesize
\begin{center}
\begin{tabular}{lllll}
\hline
\multirow{2}{*}{\textbf{Use Case}}	& \textbf{Input size}  	& \textbf{NN size}  	& \textbf{Memory}       & \textbf{Accuracy}             \\
					& (bits)		& (neurons)		& (KBytes)		& (\%)		\\			
\hline
\textit{Traffic} 	& \multirow{ 2}{*}{256} & \multirow{ 2}{*}{32, 16, 2} 	& \multirow{ 2}{*}{1.1}	& \multirow{ 2}{*}{88.6}	\\
\textit{Classification} &			&				&			&				\\
\textit{Anomaly}	& \multirow{ 2}{*}{256} & \multirow{ 2}{*}{32, 16, 2}	& \multirow{ 2}{*}{1.1}	& \multirow{ 2}{*}{85.3}	\\
\textit{Detection}	&			&				&			&				\\
\textit{Network}	& \multirow{ 2}{*}{152} & \multirow{ 2}{*}{128, 64, 2} 	& \multirow{ 2}{*}{3.4}	& \multirow{ 2}{*}{92.0}	\\
\textit{Tomography}	&			&				&			&				\\
\hline
\end{tabular}
\end{center}}
\vspace{-0.6cm}
\caption{Use cases implemented with \tonic.}
\label{tab:use-cases}
\vspace{-0.3cm}
\end{table}

We tested the use case simulating a CLOS-like Fat Tree datacenter network with ns3~\cite{ns3}, using different 
link speeds (from 100Mb/s to 10Gb/s) and traffic workloads. 
Following the methodology suggested by~\cite{geng19}, we split the problem of inferring queue sizes in 
multiple sub-problems, each targeting a subset of the queues. This allows us to run smaller MLPs on each 
of the NICs.~\footnote{This also requires a careful selection of the probe packets destinations, to guarantee 
that the probes sent to a given NIC are traversing the queues whose size is inferred by such NIC.} Unlike 
SIMON, our approach does not infer the actual size of a queue, but it only infers which queues are bigger 
than given thresholds levels. This information is usually sufficient for the control plane to take a flow-steering 
decision, while more accurate inferences, e.g., for debugging, can be still run offline
(See Figure~\ref{fig:usecase_simon}).

Each binarized MLP running with \tonic has 19 probes' one-way delays as input, and 128, 64, 2 
neurons. A NIC can run multiple of these NNs, since each of them infers the congestion status of a specific queue.
Across all the queues of the simulated network, we achieve a median accuracy in predicting a congested queue above 92\%.

\vspace{-0.1in}
\section{Evaluation}
\label{sec:evaluation}
In this section we present the experimental evaluations of \tonic. We report and discuss the 
end-to-end performance of the use cases presented in \S~\ref{sec:use_cases}, and evaluate 
the scalability of the three implementations with targeted micro-benchmarks. More details 
alongside the performance of alternative design choices are discussed in appendices.

\vspace{0.1in}
\noindent\textbf{Testbed}. 
Unless stated otherwise, we run all the tests on a machine equipped with an Intel Haswell 
E5-1630 v3 CPU and either a single Netronome Agilio CX SmartNIC, with an NFP4000 processor, 
or a NetFPGA-SUME. \footnote{We chose a Haswell CPU since it was produced with a 22nm factory process, 
i.e., a technology fairly comparable to the NFP4000 and NetFPGA Virtex7, which were produced 
with 22nm and 28nm factory processes, respectively.} The Haswell is clocked at 3.7GHz, the 
NFP at 800MHz, and the NetFPGA at 200MHz for both the \tonic-FPGA and \tonic-P4 implementations. 
The NIC under test is connected back-to-back to a 40Gb/s capable DPDK packet generator\footnote{\url{https://git.dpdk.org/apps/pktgen-dpdk/}}.
The host system runs Linux, kernel v.4.18.15. 

\begin{figure}[t!]
\begin{minipage}[t]{0.48\linewidth}
         \includegraphics[width=\linewidth]{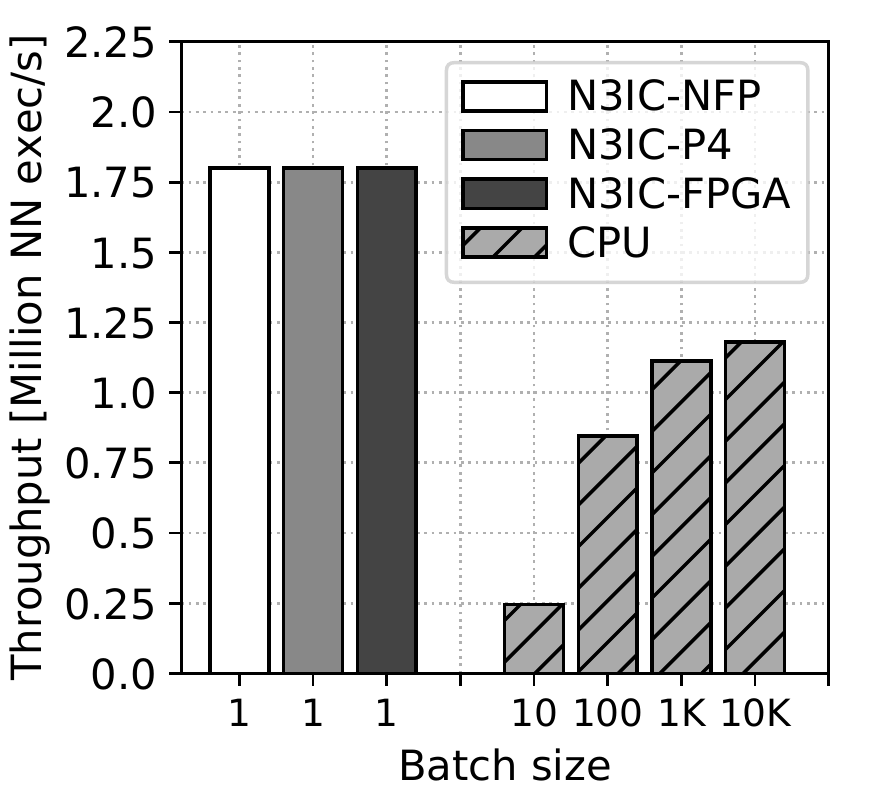}
    \vspace{-0.8cm}
         \caption{For traffic analysis, \tonic implementations match the offered load of 1.8M inferences per second, while forwarding packets at 40Gb/s. This is 1.5x the maximum throughput provided by \bnne.}
         \label{fig:tonic_tput}
    \vspace{-0.3cm}
\end{minipage}%
    \hfill%
\begin{minipage}[t]{0.48\linewidth}
         \includegraphics[width=\linewidth]{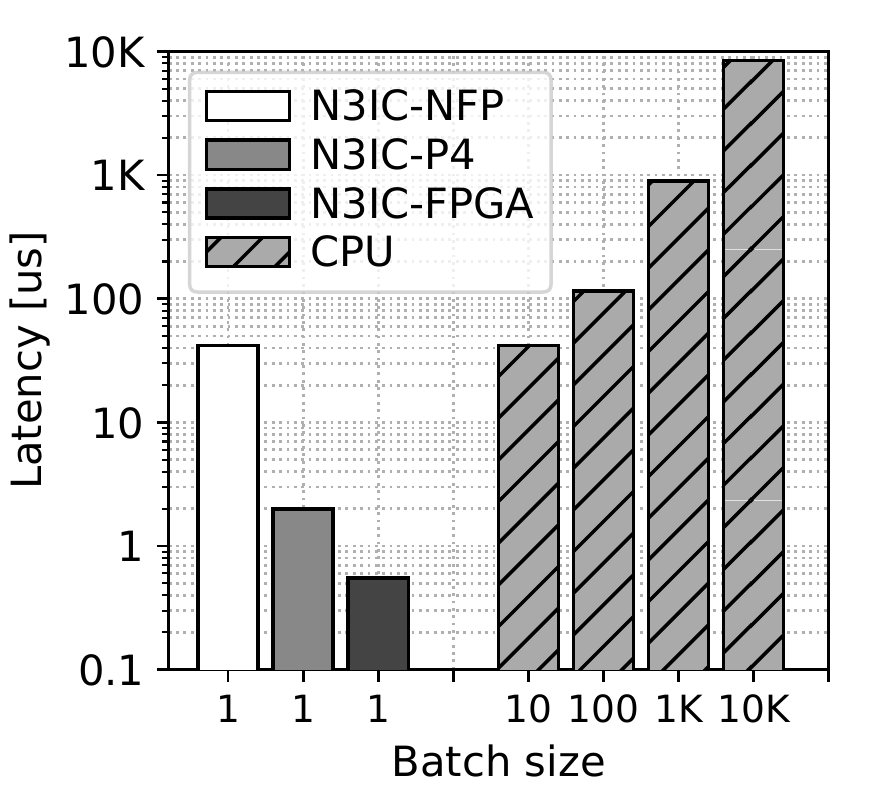}
    \vspace{-0.8cm}
         \caption{For traffic analysis use cases, \tonic implementations can provide at least 10-100x lower latency than \bnne, avoiding the need of performing batching to amortize data transfer costs. }
         \label{fig:tonic_latency}
    \vspace{-0.3cm}
\end{minipage}
\end{figure}

\vspace{0.1in}
\noindent\textbf{Comparison term}.
We compared our prototypes with a software implementation that runs binary layers (\bnne), 
available at~\cite{toNIC}. \bnne has been written in C, and optimized for the Haswell CPU, 
with some parts in assembler to take fully advantage of the CPU's architecture features, 
such as AVX instructions. We setup \bnne to read flows statistics/data from the Netronome NIC 
and ran \bnne only with the Netronome NIC since its driver is more mature than the NetFPGA's: 
it can better handle fast communication between the NIC and the host system. 
When performing inference with \bnne we took into account (1) the time to read one or 
more flow statistics; (2) the time to run the BNN itself; and (3) the time to write back 
the result on the NIC. This allows us to perform a fair comparison against \tonic.

\subsection{Traffic analysis use cases}
In both the traffic classification and anomaly detection use cases, we configured the 
NICs to collect flow statistics. We assumed that the provided traffic contains 1.8M flows 
per second, which need to be analyzed by running NN inference\footnote{That is, an average of 10 packets per flow at 40Gb/s@256B.}.
This is a challenging load for a single server, 
being more common in ToR switches handling traffic for high throughput user-facing services~\cite{silkroad}. 
Here, we observe that if \tonic can meet this performance goal, it is likely to be capable 
of handling a large range of ordinary use cases.

We first measured the NIC performance when only collecting flow statistics, that requires for each received packet:
packet parsing; a lookup in a hash-table for retrieving the flow counters; and updating several counters.
The Netronome provides 
its 40Gb/s line rate only with packets of size 256B (18.1Mpps) or bigger. This is achieved 
using 90 out of the 480 available threads, and it is inline with the device's expected performance 
for such class of applications. 
The NetFPGA, instead, is capable of forwarding 40Gb/s with minimum size (64B) packets while collecting flow statistics.

We summarized the throughput results in Figure~\ref{fig:tonic_tput}. The \tonic implementations 
can all achieve the offered throughput of 1.81M flow analysis/s, while forwarding packets at 40Gb/s 
(40Gb/s@256B in the case of the Netronome). That is, \tonic does not reduce the packet forwarding 
performance of the NICs. In comparison, even if using larger batch sizes, \bnne is unable to 
cope with such load, when running on a single CPU core. \bnne maximum throughput is 1.18M analyzed 
flows/s, when using very large batches of 10K flows. More interestingly, Figure~\ref{fig:tonic_latency} 
shows that \tonic implementations provide also a low processing latency, with a 95-th percentile 
of 42\si{\micro\second} for \tonic-NFP, and only 2\si{\micro\second} and 0.5\si{\micro\second} for \tonic-P4 
and \tonic-FPGA, respectively. In comparison, for \bnne to achieve a throughput above the 1M flows/s, 
the processing latency is 1ms and 8ms with batch sizes 1K and 10K, respectively.

\begin{tcolorbox}
\textbf{Result 1}: when performing traffic analysis, \tonic saves at least an entire CPU core, while 
providing a 1.5x higher throughput and 10-100x lower processing latency than a system running on the 
host's CPU.
\end{tcolorbox}

\begin{figure}[t!]
\begin{minipage}[t]{0.48\linewidth}
         \includegraphics[width=\linewidth]{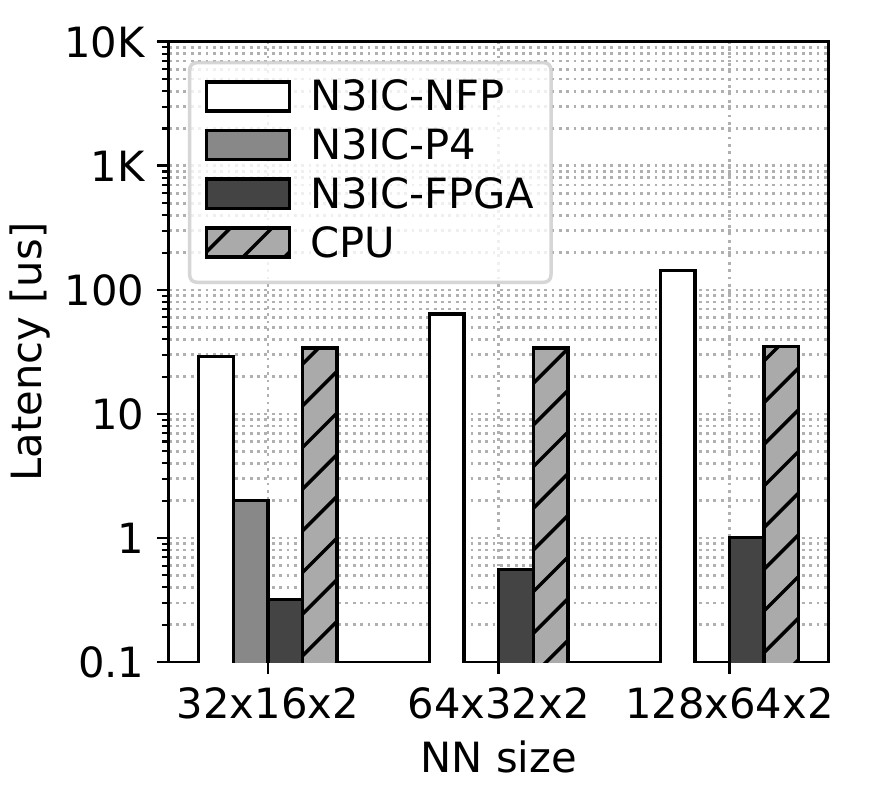}
    \vspace{-0.8cm}
         \caption{\tonic-FPGA can support the network tomography use case even in fast 400Gb/s networks with probes sent every 25\si{\micro\second}. \tonic-P4 can only run this use case with smaller and less accurate NNs.}
         \label{fig:simon_latency}
    \vspace{-0.3cm}
\end{minipage}%
    \hfill%
\begin{minipage}[t]{0.48\linewidth}
         \includegraphics[width=\linewidth]{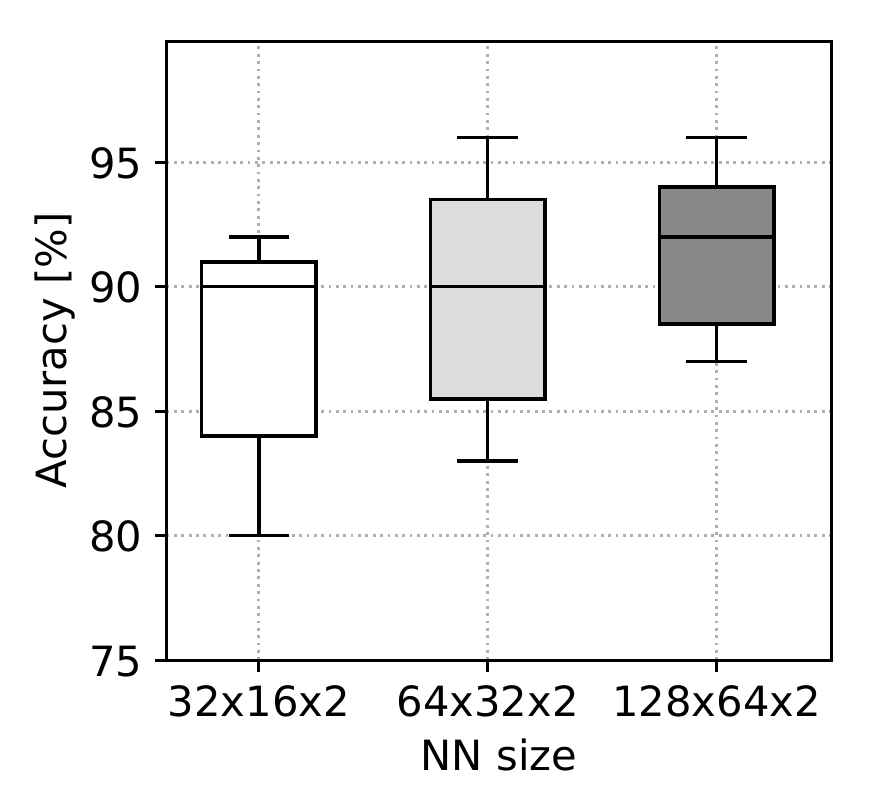}
    \vspace{-0.8cm}
         \caption{Box plot of the accuracies for the predicted queues in the network tomography use case. Larger NNs have longer execution latency but better accuracy.}
         \label{fig:simon_accuracy}
    \vspace{-0.3cm}
\end{minipage}
\end{figure}

\subsection{Network Tomography}
When testing the network tomography use case, the NIC stores the one-way-delay value for the received 
network probes, before passing them to the analysis engine, i.e., either \tonic or \bnne. Here, 
processing latency is the critical performance indicator, since in networks with 40, 100, 400Gb/s links, 
SIMON requires a new set of probe packets to be generated every 250, 100, 25\si{\micro\second}, 
respectively~\cite{geng19}.

Figure~\ref{fig:simon_latency} shows that \bnne provides a processing latency of about 40\si{\micro\second}, 
which is within the budget of 100\si{\micro\second}.\footnote{We can use a batch size of 1 in this use case, since
high-throughput is not required.} However, upcoming network links of 400Gb/s could not be supported, since 
they would lower the periodicity of the probes to 25\si{\micro\second}. \tonic processing latency for SIMON's 
NN with 128, 64, 2 neurons is 170\si{\micro\second} for \tonic-NFP and below 2\si{\micro\second} for \tonic-FPGA. 
As we further clarify next, \tonic-P4 cannot scale to run such NN, and can only run the smaller 32, 16, 2 neurons 
networks with about 2\si{\micro\second} of delay, at the cost of reduced accuracy. In Figure~\ref{fig:simon_accuracy}, we show the accuracies for the predicted queues, using different network sizes. Larger networks improve accuracy by up to 10 percentage points. Thus, for future high-performance
networks, unless a lower accuracy with a smaller NN is tolerable, only \tonic-FPGA can meet the SIMON's timing 
constraints.

\begin{tcolorbox}
\textbf{Result 2}: compared to a host-based system, for low throughput but latency-sensitive use cases \tonic-FPGA 
can reduce processing latency by 20x. This enables applications like SIMON to run in real time in networks running 
at 400Gb/s and beyond.
\end{tcolorbox}

\begin{figure}[t!]
\begin{minipage}[t]{0.48\linewidth}
         \includegraphics[width=\linewidth]{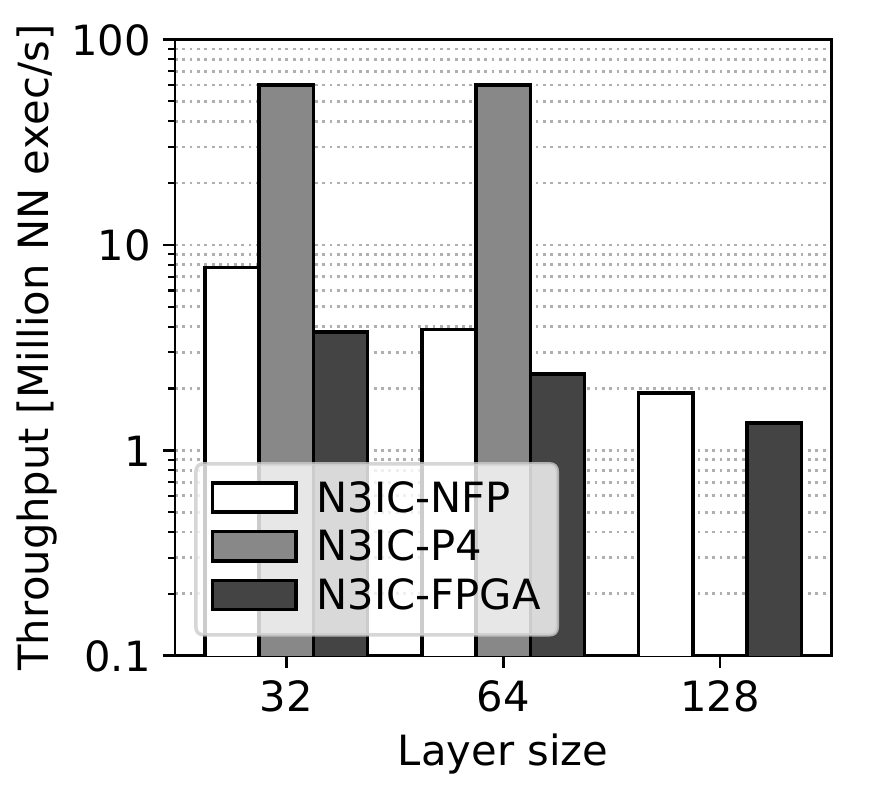}
    \vspace{-0.8cm}
         \caption{NNs execution throughput decreases linearly with the NN size. \tonic-P4 cannot scale to larger networks}
         \label{fig:scaling_tput}
    \vspace{-0.3cm}
\end{minipage}%
    \hfill%
\begin{minipage}[t]{0.48\linewidth}
         \includegraphics[width=\linewidth]{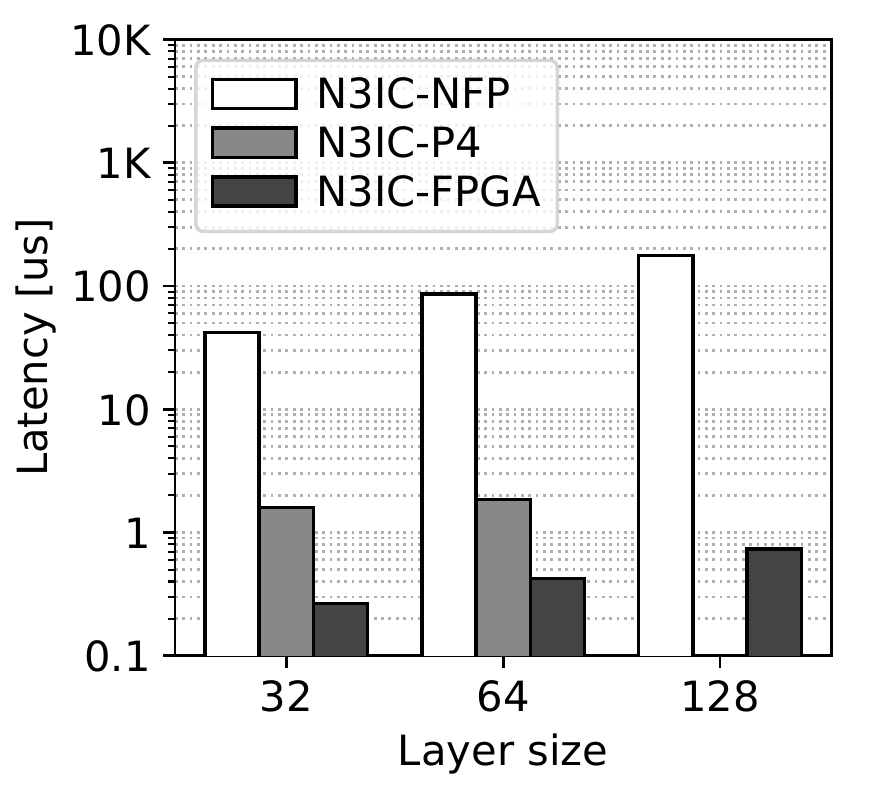}
    \vspace{-0.8cm}
         \caption{NNs execution latency increases linearly with the NN size. \tonic-P4 cannot scale to larger networks}
         \label{fig:scaling_latency}
    \vspace{-0.3cm}
\end{minipage}
\end{figure}

\subsection{Neural Network size}
We now evaluate the processing throughput and latency when varying the size of the binary neural network. We performed 
this evaluation fully loading \tonic, and by executing a single FC layer with 256 binary inputs. We varied the number 
of neurons to be 32, 64 and 128.\footnote{Notice that the FC size is number of input times number of neurons: the FC 
layer with 128 neurons has 4KB of weights, i.e., 4x the size of the NN used for the traffic analysis use cases.}

Figure~\ref{fig:scaling_tput} and Figure~\ref{fig:scaling_latency} show that both metrics have a linear decrease/increase 
with the NN size for \tonic-NFP and \tonic-FPGA, which is expected. In comparison, \tonic-P4 throughput results are 
much higher for an FC with 32 and 64 neurons. Unfortunately, results for 128 neurons are missing. As anticipated, \tonic-P4 
could not scale to handle such layers. Both results can be explained with the constraints imposed by the PISA architecture 
as implemented by the P4-NetFPGA toolchain. The computations of the NN described by P4 are expanded and unrolled in 
the FPGA, to serialize the execution in the pipelined model implemented by P4-NetFPGA. This consumes a large amount 
of the FPGA resources, thereby providing very high throughput at the cost of limited scalability. 

\begin{tcolorbox}
\textbf{Result 3}: \tonic-NFP's and \tonic-FPGA's processing throughput decreases linearly, while latency increases 
linearly, with the NN size. \tonic-P4 does not scale to larger NNs.
\end{tcolorbox}

\vspace{-0.1in}
\subsection{NIC resources usage}
Finally, we quantify the NIC resources needed by \tonic prototypes to run the NNs used in the traffic analysis 
use cases.

In the NFP case, \tonic has to store the NN's weights in the NFP4000's memory system. The NNs used with 
the traffic analysis use cases require 1.5\% of the CLS memory, and 480 threads to face the offered load, 
instead of the 90 required to achieve line rate throughput when the NIC is only collecting flow statistics. 
Here, it should be noted that it is possible to use less threads, as well as the larger and slower EMEM 
memories to store NN's weights, if a performance drop in NN inference throughput is acceptable. For instance, 
using only 120 threads, i.e., 30 additional thread compared to the baseline, reduces the throughput of flows 
analyzed per second by 10x (more details in the Appendix). This still provides the ability to analyze over 100k flows per second, which 
is sufficient for many workloads.

In the NetFPGA cases, we measured the hardware resources required to synthesize \tonic implementations 
on the Virtex7 FPGA, and compare them to the standard NetFPGA reference NIC design's resources. Table~\ref{tab:resources} 
summarizes the results. \tonic-FPGA requires only an additional 0.6\% and 1.2\% of the FPGA's LUTs and 
BRAMs, respectively. The \tonic-P4 implementation, as anticipated, requires a more significant amount 
of resources, with an additional 22\% for both LUTs and BRAMs, when compared to the reference NIC.

It is worth noticing that \tonic-FPGA is designed to match the required performance for the use cases, 
using minimum resources. The design allows for computing neuron results serially in a loop structure. 
E.g., we are able to match the throughput of \tonic-P4 for a FC layer of 32 neurons by using 
16 NN Executor modules in parallel within \tonic-FPGA. The \tonic-FPGA's throughout scales linearly with 
the number of modules, but so does the resources usage, which grows to additional 10\% of the LUTs and 
19\% of the BRAMs.\footnote{We did not optimize the NN Executor modules to share the use of CAMs across 
modules. It is thus likely possible to considerably reduce this BRAM utilization if needed.}

\begin{tcolorbox}
\textbf{Result 4}: \tonic-FPGA consumes a very small fraction of the NetFPGA resources, and both \tonic-FPGA 
and \tonic-NFP can be configured to trade-off used NICs' resources with provided performance. 
\end{tcolorbox}

\begin{table}
        \caption{NetFPGA resources usage. \tonic-FPGA requires little additional resources. \tonic-P4 uses a large amount of NIC resources due to the PISA computation model constraints.}
        \label{tab:resources}
				\vspace{-0.1in}
        \begin{center}
          \begin{small}
            \begin{sc}
              \begin{tabular}{|c|c|c|}
                \toprule
								\textbf{Design} & \textbf{LUT (\#, \% tot)} & \textbf{BRAM (\#, \% tot)} \\
                \hline
                Reference NIC &       49.4k, 11.4\%  & 194, 13.2\%  \\
								\tonic-FGPA      &       52.0k, 12.0\%  & 211, 14.4\%   \\
                \tonic-P4        &      144.5k, 33.4\% & 518, 35.2\%   \\
                \bottomrule
              \end{tabular}
            \end{sc}
          \end{small}
        \end{center}
        \vspace{-0.3cm}
\end{table}

\section{Discussion}
\label{sec:nextsteps}
Our evaluation results show that \tonic can benefit a number of networking use cases. However, like in similar solutions that leverage in-network computation~\cite{silkroad,kim18,li17,SwitchPaxos,switchKV,sapio19,busse19}, the benefits of bringing the inference closer to the wire come at a cost: (1) there
is a need to convert the machine learning models in binary NNs; and (2) applications 
have to be designed taking into account the network-host split.

A second interesting observation is that it is possible to readily implement \tonic on 
commercial programmable NICs, using existing programming models such as microC and P4, 
even though these implementations come with some limitations. \tonic-NFP is unable to 
meet latency requirements in the order of the \si{\micro\second}, while \tonic-P4 is inefficient in using the hardware resources, which significantly limits the NNs maximum size. Both limitations can
be overcame if the NIC supports natively binary NN execution as a primitive with 
a dedicated hardware module, as shown by \tonic-FPGA. In fact, our tests show that supporting 
such primitive comes at very little overhead in terms of required hardware resources.

Furthermore, \tonic-FPGA can be considerably more efficient in running the NNs, thereby
enabling larger NNs, or even supporting multiple different NNs at the same time. 
For instance, in the network tomography use case, our modified version of SIMON runs several
small NNs, each predicting the congestion status of a given queue.
In \tonic-NFP, running them requires a good share of the network processor's threads
to compute the NNs in parallel, leaving little space for other computations.
Instead, \tonic-FPGA uses a single NN executor module, which serially processes
NNs one after the other, while still respecting the strict timing constraints of the use case.
Each NN execution takes in fact few \si{\micro\second}, and given the small resource overhead of the 
NN executor module, it would be possible to include multiple of them if the need arises.

Finally, it should be noted that \tonic is not a generic tool to run any NN on 
the NIC. For large NNs, e.g., with thousands of neurons, 
its execution time may grow significantly, making the overhead of the PCIe transfer relatively 
lower. In such cases, a better choice is the use of a different NN executor, i.e., 
BrainWave and Taurus, or keeping the computation on the host system. More details in the Appendix.

\section{Related Work}
\label{sec:related}
\tonic relates with the works that move machine learning capabilities into programmable 
hardware. This has been explored in the context of programmable switches~\cite{busse19,xiong19}, 
or by using fully-fledged network-attached accelerators~\cite{brainwave,swamy19}. In this 
paper, instead, we show that it is possible to design and implement a significantly 
smaller scale NN executor in the data plane of commodity off-the-shelf programmable NICs. 

In this direction, we extend some ideas presented in~\cite{n2net, Sanvito}. However, those 
works focused either on a conceptual design for RMT~\cite{forwardingMetamorphosis} switches~\cite{n2net}, 
or on end-host ML applications, in which the NIC works as a co-processor for CNNs running on 
the host. Instead, in this paper, we present a complete evaluation of BNN executors on two 
NICs, propose a dedicated hardware-native implementation, and include an end-to-end evaluation 
of three networking use cases, with related trade-offs in terms of model size and specific 
hardware limitations. \tonic can be also positioned in the larger trend of in-network computing 
research. In fact, researchers have been exploring ways to apply programmable switches to 
domains not necessarily related to networking, such as key-value store~\cite{switchKV}, 
distributed consensus~\cite{SwitchPaxos}, and the acceleration of parameter servers for ML 
model training~\cite{daiet, sapio19}. Another related area is the design of hardware for NN 
accelerators.
A valuable survey on the topic is provided by~\cite{surveyHwDNN}. Particularly relevant to 
our work are architectures such as YodaNN~\cite{yodaNN} and FINN~\cite{finn}. 
Finally, while not directly related to \tonic, recent work on the security of network
applications that use machine learning~\cite{VanbeverHotnets19} is likely 
to influence developments in this area.

\section{Conclusion}
\label{sec:conclusion}

In this paper, we presented \tonic, a technique to run neural networks in the data 
plane of commodity programmable NICs. We experimentally evaluated \tonic on the 
Netronome NFP4000 and on the NetFPGA. We showed that \tonic can greatly benefit 
network applications discussing three use cases: traffic classification, anomaly 
detection and network tomography. Our results show that \tonic can be readily 
implemented in existing commercial programmable NICs. By doing so, in the traffic 
analysis use cases, we demonstrated that \tonic provides 1.5x higher throughput 
and 100x lower latency than a state-of-the-art software implementation, while 
saving precious CPU resources. At the same time, supporting \tonic as a primitive 
with dedicated hardware requires only little additional hardware resources: 
the \tonic NetFPGA implementation increases the logic and memory resources of a 
standard NIC design by less than 2\%, and it enables a real time network tomography 
use case, which would be otherwise unfeasible.

\bibliographystyle{plain}
\bibliography{biblio}

\begin{thebibliography}{10}

\bibitem{yodaNN}
Renzo Andri, Lukas Cavigelli, Davide Rossi, and Luca Benini.
\newblock {YodaNN: an ultra-low power convolutional neural network accelerator
  based on binary weights}.
\newblock In {\em Annual Symposium on VLSI (ISVLSI)}. IEEE, 2016.

\bibitem{ballani15}
Hitesh Ballani, Paolo Costa, Christos Gkantsidis, Matthew~P. Grosvenor, Thomas
  Karagiannis, Lazaros Koromilas, and Greg O'Shea.
\newblock {Enabling End-Host Network Functions}.
\newblock In {\em Special Interest Group on Data Communication (SIGCOMM)}. ACM,
  2015.

\bibitem{barbette15}
Tom Barbette, Cyril Soldani, and Laurent Mathy.
\newblock {Fast Userspace Packet Processing}.
\newblock In {\em Architectures for Networking and Communications Systems
  (ANCS)}. IEEE/ACM, 2015.

\bibitem{beeler1972hakmem}
M.~Beeler, R.W. Gosper, and R.~Schroeppel.
\newblock {Hakmem AI Memo No. 239}.
\newblock In {\em MIT Artificial Intelligence Laboratory, Cambridge, US}, 1972.

\bibitem{p4}
Pat Bosshart, Dan Daly, Glen Gibb, Martin Izzard, Nick McKeown, Jennifer
  Rexford, Cole Schlesinger, Dan Talayco, Amin Vahdat, George Varghese, and
  David Walker.
\newblock {P4: Programming Protocol-independent Packet Processors}.
\newblock In {\em Computer Communication Review, Volume: 44, Issue: 3}. ACM,
  2014.

\bibitem{forwardingMetamorphosis}
Pat Bosshart, Glen Gibb, Hun-Seok Kim, George Varghese, Nick McKeown, Martin
  Izzard, Fernando Mujica, and Mark Horowitz.
\newblock {Forwarding Metamorphosis: Fast Programmable Match-action Processing
  in Hardware for SDN}.
\newblock In {\em Special Interest Group on Data Communication (SIGCOMM)}. ACM,
  2013.

\bibitem{bujlow2015independent}
Tomasz Bujlow, Valent{\'\i}n Carela-Espa{\~n}ol, and Pere Barlet-Ros.
\newblock {Independent comparison of popular DPI tools for traffic
  classification}.
\newblock In {\em Computer Networks, Volume: 76, Issue: C}. Elsevier, 2015.

\bibitem{busse19}
Coralie Busse-Grawitz, Roland Meier, Alexander Dietmuller, Tobiad Buhler, and
  Laurent Vanbever.
\newblock {pForest: In-Network Inference with Random Forests}.
\newblock In {\em Computing Research Repository, Volume: abs/1909.05680}, 2019.

\bibitem{CaiHSV17}
Zhaowei Cai, Xiaodong He, Jian Sun, and Nuno Vasconcelos.
\newblock {Deep Learning with Low Precision by Half-wave Gaussian
  Quantization}.
\newblock In {\em Computing Research Repository, Volume: abs/1702.00953}, 2017.

\bibitem{trainingBNN}
Matthieu Courbariaux, Itay Hubara, Daniel Soudry, Ran El-Yaniv, and Yoshuao
  Bengio.
\newblock Binarized neural networks: Training deep neural networks with weights
  and activations constrained to+ 1 or-1.
\newblock In {\em Computing Research Repository, Volume: abs/1602.02830}, 2016.

\bibitem{SwitchPaxos}
Huynh~Tu Dang, Marco Canini, Fernando Pedone, and Robert Soul{\'e}.
\newblock {Paxos made switch-y}.
\newblock In {\em Computer Communication Review, Volume: 46, Issue: 2}. ACM,
  2016.

\bibitem{eran19}
Haggai Eran, Lior Zeno, Maroun Tork, Gabi Malka, and Mark Silberstein.
\newblock {NICA: An Infrastructure for Inline Acceleration of Network
  Applications}.
\newblock In {\em Annual Technical Conference (ATC)}. USENIX Association, 2019.

\bibitem{esmaeilzadeh11}
Hadi Esmaeilzadeh, Emily Blem, Renee St.~Amant, Karthikeyan Sankaralingam, and
  Doug Burger.
\newblock {Dark silicon and the end of multicore scaling}.
\newblock In {\em International Symposium on Computer Architecture (ISCA)}.
  ACM, 2011.

\bibitem{feamster18}
Nick Feamster and Jennifer Rexford.
\newblock {Why (and How) Networks Should Run Themselves}.
\newblock In {\em Applied Networking Research Workshop (ANRW)}. ACM, 2018.

\bibitem{firestone17}
Daniel Firestone.
\newblock {Building hardware-accelerated networks at scale in the c/guloud},
  Jan 2019.
\newblock
  \url{https://conferences.sigcomm.org/sigcomm/2017/files/program-kbnets/keynote-2.pdf}.

\bibitem{firestone18}
Daniel Firestone, Andrew Putnam, Sambhrama Mundkur, Derek Chiou, Alireza
  Dabagh, Mark Andrewartha, Hari Angepat, Vivek Bhanu, Adrian Caulfield, Eric
  Chung, Harish~K. Chandrappa, Somesh Chaturmohta, Matt Humphrey, Jack Lavier,
  Norman Lam, Fengfen Liu, Kalin Ovtcharov, Jitu Padhye, Gautham Popuri,
  Shachar Raindel, Tejas Sapre, Mark Shaw, Gabriel Silva, Madhan Sivakumar,
  Nisheeth Srivastava, Anshuman Verma, Qasim Zuhair, Deepak Bansal, Doug
  Burger, Kushagra Vaid, David~A. Maltz, and Albert Greenberg.
\newblock {Azure Accelerated Networking: SmartNICs in the Public Cloud}.
\newblock In {\em Networked Systems Design and Implementation (NSDI)}. USENIX,
  2018.

\bibitem{forman2003extensive}
George Forman.
\newblock {An extensive empirical study of feature selection metrics for text
  classification}.
\newblock In {\em Journal of machine learning research, Volume: 3}. JMLR.org,
  2003.

\bibitem{geng19}
Yilong Geng, Shiyu Liu, Zi~Yin, Ashish Naik, Balaji Prabhakar, Mendel
  Rosenblum, and Amin Vahdat.
\newblock {SIMON: A Simple and Scalable Method for Sensing, Inference and
  Measurement in Data Center Networks}.
\newblock In {\em Networked Systems Design and Implementation (NSDI)}. USENIX,
  2019.

\bibitem{ghasemi17}
Mojgan Ghasemi, Theophilus Benson, and Jennifer Rexford.
\newblock {Dapper: Data Plane Performance Diagnosis of TCP}.
\newblock In {\em Symposium on SDN Research (SOSR)}. ACM, 2017.

\bibitem{gupta15}
Prabhat~K. Gupta.
\newblock {Xeon+FPGA Platform for the Data Center}, 2019.
\newblock
  \url{https://www.ece.cmu.edu/~calcm/carl/lib/exe/fetch.php?media=carl15-gupta.pdf}.

\bibitem{Han2015DeepCC}
Song Han, Huizi Mao, and William~J. Dally.
\newblock {Deep Compression: Compressing Deep Neural Network with Pruning,
  Trained Quantization and Huffman Coding}.
\newblock In {\em Computing Research Repository, Volume: abs/1510.00149}, 2015.

\bibitem{theRiseOfDarkSilicon}
Nikos Hardavellas.
\newblock {The rise and fall of dark silicon}.
\newblock In {\em ;login:, Volume: 37}. USENIX, 2012.

\bibitem{hardavellas11}
Nikos Hardavellas, Michael Ferdman, Babak Falsafi, and Anastasia Ailamaki.
\newblock {Toward dark silicon in servers}.
\newblock In {\em Micro, Volume: 31, Issue: 4}. IEEE, 2011.

\bibitem{hazelwood18}
Kim Hazelwood, Sarah Bird, David Books, Soumith Chintala, Utku Diril, Dmytro
  Dzhulgakov, Mohamed Fawzy, Bill Jia, Yangqing Jia, Aditya Kalro, James Law,
  Kevin Lee, Jason Lu, Pieter Noordhuis, Misha Smelyanskiy, Liang Xiong, and
  Xiaodong Wang.
\newblock {Applied Machine Learning at Facebook: A Datacenter Infrastructure
  Perspective}.
\newblock In {\em High Performance Computer Architecture (HPCA)}. IEEE, 2018.

\bibitem{fbml}
Kim Hazelwood, Sarah Bird, David Brooks, Soumith Chintala, Utku Diril, Dmytro
  Dzhulgakov, Mohamed Fawzy, Bill Jia, Yangqing Jia, Aditya Kalro, James Law,
  Kevin Lee, Jason Lu, Pieter Noordhuis, Misha Smelyanskiy, Liang Xiong, and
  Xiaodong Wang.
\newblock {Applied machine learning at Facebook: a datacenter infrastructure
  perspective}.
\newblock In {\em High Performance Computer Architecture (HPCA)}. IEEE, 2018.

\bibitem{hennessy19}
John~L. Hennessy and David~A. Patterson.
\newblock {A New Golden Age for Computer Architecture}.
\newblock In {\em Communications of the ACM, Volume: 62, Issue: 2}. ACM, 2019.

\bibitem{Hornik}
Kur Hornik, Maxwell Stinchcombe, and Halber White.
\newblock {Multilayer feedforward networks are universal approximators}.
\newblock In {\em Neural Networks, Volume: 2, Issue: 5}. Elsevier Science Ltd.,
  1989.

\bibitem{bnn}
Itay Hubara, Matthieu Courbariaux, Daniel Soudry, Ran El-Yaniv, and Yoshua
  Bengio.
\newblock {Binarized neural networks}.
\newblock In {\em Neural Information Processing Systems (NIPS)}. Curran
  Associates Inc., 2016.

\bibitem{ibanez19}
Stephen Ibanez, Gordon Brebner, Nick McKeown, and Noa Zilberman.
\newblock The p4-netfpga workflow for line-rate packet processing.
\newblock In {\em Field-Programmable Gate Arrays (FPGA)}. ACM, 2019.

\bibitem{tpu}
Norman~P. Jouppi, Cliff Young, Nishant Patil, David Patterson, Gaurav Agrawal,
  Raminder Bajwa, Sarah Bates, Suresh Bhatia, Nan Boden, Al~Borchers, Rick
  Boyle, Pierre-luc Cantin, Clifford Chao, Chris Clark, Jeremy Coriell, Mike
  Daley, Matt Dau, Jeffrey Dean, Ben Gelb, Tara~Vazir Ghaemmaghami, Rajendra
  Gottipati, William Gulland, Robert Hagmann, C.~Richard Ho, Doug Hogberg, John
  Hu, Robert Hundt, Dan Hurt, Julian Ibarz, Aaron Jaffey, Alek Jaworski,
  Alexander Kaplan, Harshit Khaitan, Daniel Killebrew, Andy Koch, Naveen Kumar,
  Steve Lacy, James Laudon, James Law, Diemthu Le, Chris Leary, Zhuyuan Liu,
  Kyle Lucke, Alan Lundin, Gordon MacKean, Adriana Maggiore, Maire Mahony,
  Kieran Miller, Rahul Nagarajan, Ravi Narayanaswami, Ray Ni, Kathy Nix, Thomas
  Norrie, Mark Omernick, Narayana Penukonda, Andy Phelps, Jonathan Ross, Matt
  Ross, Amir Salek, Emad Samadiani, Chris Severn, Gregory Sizikov, Matthew
  Snelham, Jed Souter, Dan Steinberg, Andy Swing, Mercedes Tan, Gregory
  Thorson, Bo~Tian, Horia Toma, Erick Tuttle, Vijay Vasudevan, Richard Walter,
  Walter Wang, Eric Wilcox, and Doe~Hyun Yoon.
\newblock {In-datacenter performance analysis of a Tensor Processing Unit}.
\newblock In {\em International Symposium on Computer Architecture (ISCA)}.
  ACM, 2017.

\bibitem{karagiannis05}
Thomas Karagiannis, Konstantina Papagiannaki, and Michalis Faloutsos.
\newblock {BLINC: Multilevel Traffic Classification in the Dark}.
\newblock In {\em Special Interest Group on Data Communication (SIGCOMM)}. ACM,
  2005.

\bibitem{intelcaffe}
Vadim Karpusenko, Andres Rodriguez, Jacek Czaja, and Mariusz Moczala.
\newblock {Caffe* optimized for Intel architecture: applying modern code
  techniques}.
\newblock In {\em Technical Report}. Intel, 2016.

\bibitem{kathareios17}
Georgios Kathareios, Andreea Anghel, Akos Mate, Rolf Clauberg, and Mitch Gusat.
\newblock {Catch It If You Can: Real-Time Network Anomaly Detection with Low
  False Alarm Rates}.
\newblock In {\em International Conference on Machine Learning and Applications
  (ICMLA)}. IEEE, 2017.

\bibitem{katsikas18}
Georgios~P. Katsikas, Tom Barbette, Dejan Kostiundefined, Rebecca Steinert, and
  Gerald~Q. Maguire.
\newblock {Metron: NFV Service Chains at the True Speed of the Underlying
  Hardware}.
\newblock In {\em Networked Systems Design and Implementation (NSDI)}. USENIX
  Association, 2018.

\bibitem{kim18}
Daehyeok Kim, Amirsaman Memaripour, Anirudh Badam, Yibo Zhu, Hongqiang~Harry
  Liu, Jitu Padhye, Shachar Raindel, Steven Swanson, Vyas Sekar, and Srinivasan
  Seshan.
\newblock {Hyperloop: Group-based NIC-offloading to Accelerate Replicated
  Transactions in Multi-tenant Storage Systems}.
\newblock In {\em Special Interest Group on Data Communication (SIGCOMM)}. ACM,
  2018.

\bibitem{kraska19}
Tim Kraska, Mohammad Alizadeh, Alex Beutel, Ed~H. Chi, Jialin Ding, Ani Kristo,
  Guillaume Leclerc, Samuel Madden, Hongzi Mao, and Vikram Nathan.
\newblock {SageDB: a learned database system}.
\newblock In {\em Conference on Innovative Data Systems Research (CIDR)}, 2019.

\bibitem{kraska18}
Tim Kraska, Alex Beutel, Ed~H. Chi, Jeffrey Dean, and Neoklis Polyzotis.
\newblock {The case for learned index structures}.
\newblock In {\em Conference on Management of Data (SIGMOD)}. ACM, 2018.

\bibitem{Lee_2017}
Edward~H. Lee, Daisuke Miyashita, Elaina Chai, Boris Murmann, and Simon~S.
  Wong.
\newblock {LogNet: Energy-efficient neural networks using logarithmic
  computation}.
\newblock In {\em International Conference on Acoustics, Speech and Signal
  Processing (ICASSP)}. IEEE, 2017.

\bibitem{li17}
Bojie Li, Zhenyuan Ruan, Wencong Xiao, Yuanwei Lu, Yongqiang Xiong, Andrew
  Putnam, Enhong Chen, and Lintao Zhang.
\newblock {KV-Direct: High-Performance In-Memory Key-Value Store with
  Programmable NIC}.
\newblock In {\em Symposium on Operating Systems Principles (SOSP)}. ACM, 2017.

\bibitem{Li2016TernaryWN}
Fengfu Li and Bin Liu.
\newblock {Ternary Weight Networks}.
\newblock In {\em Computing Research Repository, Volume: abs/1605.04711}, 2016.

\bibitem{switchKV}
Xiaozhou Li, Raghav Sethi, Michael Kaminsky, David~G. Andersen, and Michael~J.
  Freedman.
\newblock Be fast, cheap and in control with switchkv.
\newblock In {\em Networked Systems Design and Implementation (NSDI)}. USENIX,
  2016.

\bibitem{liang19}
Eric Liang, Hang Zhu, Xin Jin, and Ion Stoica.
\newblock {Neural Packet Classification}.
\newblock In {\em Special Interest Group on Data Communication (SIGCOMM)}. ACM,
  2019.

\bibitem{BNN_NIPS2017}
Xiaofan Lin, Cong Zhao, and Wei Pan.
\newblock {Towards accurate binary convolutional neural network}.
\newblock In {\em Neural Information Processing Systems (NIPS)}. Curran
  Associates, Inc., 2017.

\bibitem{luo18}
Layong Luo.
\newblock {Towards converged smartNIC architecture for bare metal \& public
  clouds}, 2018.
\newblock
  \url{https://conferences.sigcomm.org/events/apnet2018/slides/larry.pdf}.

\bibitem{martins13}
Joao Martins, Mohamed Ahmed, Costin Raiciu, and Felipe Huici.
\newblock {Enabling Fast, Dynamic Network Processing with ClickOS}.
\newblock In {\em Hot Topics in Software Defined Networking (HotSDN)}. ACM,
  2013.

\bibitem{martins14}
Joao Martins, Mohamed Ahmed, Costin Raiciu, Vladimir Olteanu, Michio Honda,
  Roberto Bifulco, and Felipe Huici.
\newblock {ClickOS and the Art of Network Function Virtualization}.
\newblock In {\em Networked Systems Design and Implementation (NSDI)}. USENIX
  Association, 2014.

\bibitem{VanbeverHotnets19}
Roland Meier, Thomas Holterbach, Stephan Keck, Matthias St\"{a}hli, Vincent
  Lenders, Ankit Singla, and Laurent Vanbever.
\newblock {(Self) Driving Under the Influence: Intoxicating Adversarial Network
  Inputs}.
\newblock In {\em Hot Topics in Networks (HotNets)}. ACM, 2019.

\bibitem{silkroad}
Rui Miao, Hongyi Zeng, Changhoon Kim, Jeongkeun Lee, and Minlan Yu.
\newblock {SilkRoad: making stateful layer-4 load balancing fast and cheap
  using switching ASICs}.
\newblock In {\em Special Interest Group on Data Communication (SIGCOMM)}. ACM,
  2017.

\bibitem{brainwave}
Microsoft.
\newblock Microsoft unveils project brainwave for real-time ai, 2017.
\newblock
  \url{https://www.microsoft.com/en-us/research/blog/microsoft-unveils-project-brainwave/}.

\bibitem{moore05}
Andrew~W. Moore and Denis Zuev.
\newblock {Internet Traffic Classification Using Bayesian Analysis Techniques}.
\newblock In {\em Conference on Measurement and Modeling of Computer Systems
  (SIGMETRICS)}. ACM, 2005.

\bibitem{moshref16}
Masoud Moshref, Minlan Yu, Ramesh Govindan, and Amin Vahdat.
\newblock {Trumpet: Timely and Precise Triggers in Data Centers}.
\newblock In {\em Special Interest Group on Data Communication (SIGCOMM)}. ACM,
  2016.

\bibitem{UNSW_NB15_paper}
Nour Moustafa and Jill Slay.
\newblock {UNSW-NB15: a comprehensive data set for network intrusion detection
  systems (UNSW-NB15 network data set)}.
\newblock In {\em Military Communications and Information Systems Conference
  (MilCIS)}. IEEE, 2015.

\bibitem{netronome}
Netronome.
\newblock Netronome {AgilioTM CX 2x40GbE} intelligent server adapter, 2018.
\newblock
  \url{https://www.netronome.com/media/redactor_files/PB_Agilio_CX_2x40GbE.pdf}.

\bibitem{netronome_edge}
Netronome.
\newblock {Packet and Netronome innovate on smart networking-focused edge
  compute hardware}, December 2018.
\newblock
  \url{https://www.netronome.com/press-releases/packet-and-netronome-innovate-smart-networking-focused-edge-compute-hardware/}.

\bibitem{neugebauer18}
Rolf Neugebauer, Gianni Antichi, Jos{\'e}~Fernando Zazo, Yury Audzevich, Sergio
  L\'{o}pez-Buedo, and Andrew~W. Moore.
\newblock {Understanding PCIe Performance for End Host Networking}.
\newblock In {\em Special Interest Group on Data Communication (SIGCOMM)}. ACM,
  2018.

\bibitem{gpudirect}
Nvidia.
\newblock {GPUDirect Technology}, [Online; accessed 10-Jan-2020].
\newblock \url{https://developer.nvidia.com/gpudirect}.

\bibitem{ousterhout15}
Kay Ousterhout, Ryan Rasti, Sylvia Ratnasamy, Scott Shenker, and Byung-Gon
  Chun.
\newblock {Making Sense of Performance in Data Analytics Frameworks}.
\newblock In {\em Networked Systems Design and Implementation (NSDI)}. USENIX,
  2015.

\bibitem{toNIC}
{Paper anonymized authors}.
\newblock Paper anonymized source code repository, 2019.
\newblock \url{https://paper-anonymized-link}.

\bibitem{zeek}
Vern Paxon.
\newblock {The Zeek Network Security Monitor}, [Online; accessed 04-Feb-2020].
\newblock \url{https://www.zeek.org/}.

\bibitem{XNOR-net}
Mohammad Rastegari, Vicente OrdonezJ, Joseph Redmon, and Ali Farhadi.
\newblock {XNOR-Net: imageNet classification using binary convolutional neural
  networks}.
\newblock In {\em Lecture Notes in Computer Science, Volume: 9908}. Springer,
  Cham, 2016.

\bibitem{roy15}
Arjun Roy, Hongyi Zeng, Jasmeet Bagga, George Porter, and Alex~C. Snoeren.
\newblock {Inside the Social Network's (Datacenter) Network}.
\newblock In {\em Special Interest Group on Data Communication (SIGCOMM)}. ACM,
  2015.

\bibitem{Sanvito}
Davide Sanvito, Giuseppe Siracusano, and Roberto Bifulco.
\newblock {Can the network be the AI accelerator?}
\newblock In {\em In-Network Computing (NetCompute)}. ACM, 2018.

\bibitem{daiet}
Amedeo Sapio, Ibrahim Abdelaziz, Abdulla Aldilaijan, Marco Canini, and Panos
  Kalnis.
\newblock {In-network computation is a dumb idea whose time has come}.
\newblock In {\em Hot Topics in Networks (HotNets)}. ACM, 2017.

\bibitem{sapio19}
Amedeo Sapio, Marco Canini, Chen-Yu Ho, Jacob Nelson, Panos Kalnis, Changhoon
  Kim, Arvind Krishnamurthy, Masoud Moshref, Dan~R.K. Ports, and Peter
  Richtarik.
\newblock {Scaling Distributed Machine Learning with In-Network Aggregation}.
\newblock In {\em Computing Research Repository, Volume: abs/1903.06701}, 2019.

\bibitem{sekar12}
Vyas Sekar, Norbert Egi, Sylvia Ratnasamy, Michael~K. Reiter, and Guangyu Shi.
\newblock {Design and Implementation of a Consolidated Middlebox Architecture}.
\newblock In {\em Networked Systems Design and Implementation (NSDI)}. USENIX
  Association, 2012.

\bibitem{sherry15}
Justine Sherry, Peter~Xiang Gao, Soumya Basu, Aurojit Panda, Arvind
  Krishnamurthy, Christian Maciocco, Maziar Manesh, Jo\~{a}o Martins, Sylvia
  Ratnasamy, Luigi Rizzo, and et~al.
\newblock {Rollback-Recovery for Middleboxes}.
\newblock In {\em Special Interest Group on Data Communication (SIGCOMM)}. ACM,
  2015.

\bibitem{vgg}
Karen Simonyan and Andrew Zisserman.
\newblock {Very deep convolutional networks for large-scale image recognition}.
\newblock In {\em Computing Research Repository, Volume: abs/1409.1556}, 2014.

\bibitem{n2net}
Giuseppe Siracusano and Roberto Bifulco.
\newblock {In-network neural networks}.
\newblock In {\em Computing Research Repository, Volume: abs/1801.05731}, 2018.

\bibitem{sivaraman16}
Anirudh Sivaraman, Alvin Cheung, Mihai Budiu, Changhoon Kim, Mohammad Alizadeh,
  Hari Balakrishnan, George Varghese, Nick McKeown, and Steve Licking.
\newblock {Packet transactions: high-level programming for line-rate switches}.
\newblock In {\em Special Interest Group on Data Communication (SIGCOMM)}. ACM,
  2016.

\bibitem{swamy19}
Tushar Swamy, Alexander Rucker, Muhammad Shahbaz, Neeraja Yadwadkar, Yaqi
  Zhang, and Kunle Olukotun.
\newblock {Taurus: An Intelligent Data Plane}.
\newblock In {\em P4 Workshop}, 2019.

\bibitem{surveyHwDNN}
Vivienne Sze, Yu-Hsin Chen, Tien-Ju Yang, and Joel Emer.
\newblock {Efficient processing of deep neural networks: a tutorial and
  survey}.
\newblock In {\em Proceedings of the IEEE, Volume: 105, Issue: 12}. IEEE, 2017.

\bibitem{bluefield}
Mellanox Technologies.
\newblock {BlueField SmartNIC}, 2019.
\newblock
  \url{http://www.mellanox.com/related-docs/prod_adapter_cards/PB_BlueField_Smart_NIC.pdf}.

\bibitem{tilencore}
Mellanox Technologies.
\newblock {TILEncore-Gx72}, 2019.
\newblock
  \url{https://www.mellanox.com/page/products_dyn?product_family=231&mtag=tilencore_gx72_adapter_mtag}.

\bibitem{ns3}
{The University of Washington NS-3 Consortium}.
\newblock {NS3 official website}, [Online; accessed 10-Jan-2020].
\newblock \url{https://www.nsnam.org/}.

\bibitem{tilmans18}
Olivier Tilmans, Tobias B\"{u}hler, Ingmar Poese, Stefano Vissicchio, and
  Laurent Vanbever.
\newblock {Stroboscope: Declarative Network Monitoring on a Budget}.
\newblock In {\em Networked Systems Design and Implementation (NSDI)}. USENIX
  Association, 2018.

\bibitem{tyson17}
Gareth Tyson, Shan Huang, Felix Cuadrado, Ignacio Castro, Vasile~C Perta,
  Arjuna Sathiaseelan, and Steve Uhlig.
\newblock {Exploring HTTP header manipulation in-the-wild}.
\newblock In {\em World Wide Web (WWW)}. International World Wide Web
  Conferences Steering Committee, 2017.

\bibitem{finn}
Yaman Umuroglu, Nicholas~J. Fraser, Giulio Gambardella, Michaela Blott, Philip
  Leong, Magnus Jahre, and Kees Vissers.
\newblock {FINN: A Framework for Fast, Scalable Binarized Neural Network
  Inference}.
\newblock In {\em Field-Programmable Gate Arrays (FPGA)}. ACM, 2017.

\bibitem{xilinx19}
Xilinx.
\newblock {SDNet compiler}, 2019.
\newblock \url{https://www.xilinx.com/sdnet}.

\bibitem{xiong19}
Zhaoqi Xiong and Noa Zilberman.
\newblock {Do Switches Dream of Machine Learning? Toward In-Network
  Classification}.
\newblock In {\em Hot Topics in Networks (HotNets)}. ACM, 2019.

\bibitem{haddadi}
Chaoyun Zhang, Paul Patras, and Hamed Haddadi.
\newblock {Deep learning in mobile and wireless networking: A survey}.
\newblock In {\em Computing Research Repository, Volume: abs/1803.04311}, 2018.

\bibitem{ZhouNZWWZ16}
Shuchang Zhou, Zekun Ni, Xinyu Zhou, He~Wen, Yuxin Wu, and Yuheng Zou.
\newblock {DoReFa-Net: Training Low Bitwidth Convolutional Neural Networks with
  Low Bitwidth Gradients}.
\newblock In {\em Computing Research Repository, Volume: abs/1606.06160}, 2016.

\bibitem{Zhu2016TrainedTQ}
Chenzhuo Zhu, Song Han, Huizi Mao, and William~J. Dally.
\newblock {Trained Ternary Quantization}.
\newblock In {\em Computing Research Repository, Volume: abs/1612.01064}, 2016.

\bibitem{netsume}
Noa Zilberman, Yury Audzevich, Adam~G. Covington, and Andrew~W. Moore.
\newblock {NetFPGA SUME: toward 100 {Gbps} as research commodity}.
\newblock In {\em Micro, Volume: 34, Issue: 5}. IEEE, 2014.

\end{thebibliography}

\newpage
\clearpage
\begin{appendix}
\label{sec:appendix}

\section{Netronome: implementation details}
\label{sec:appendix_netronome}
This section reports additional detail regarding the \tonic implementation on the Netronome NFP4000 when 
running large NNs. Netronome NFP4000 NICs have several different memory areas, with different sizes and 
access performance. Table~\ref{tab:access_times} summarizes the memory properties. 

\begin{table}[h!]
        \caption{Access times and size for the different memory areas of an NFP4000.}
        \label{tab:access_times}
        \begin{center}
                \begin{small}
                        \begin{sc}
                                \begin{tabular}{|c|c|c||c|}
                                        \toprule
                                        \textbf{Memory} & \multicolumn{2}{|c||}{\textbf{Access time (ns)}} & \textbf{Memory}\\
                                        \textbf{Type} & Min &   Max     &       \textbf{Size}   \\
                                        \hline
                                        cls                     &       25      &       62.5    &       64KB    \\
                                        ctm                     &               62.5    &       125     &       256KB   \\
                                        imem            &               187.5   &       312.5   &       4MB             \\
                                        emem            &               312.5   &       625     &       3MB             \\
                                        \bottomrule
                                \end{tabular}
                        \end{sc}
                \end{small}
        \end{center}
        \vspace{-0.15in}
\end{table}

\begin{figure}[t!]
        \centering
        \includegraphics[width=.6\columnwidth]{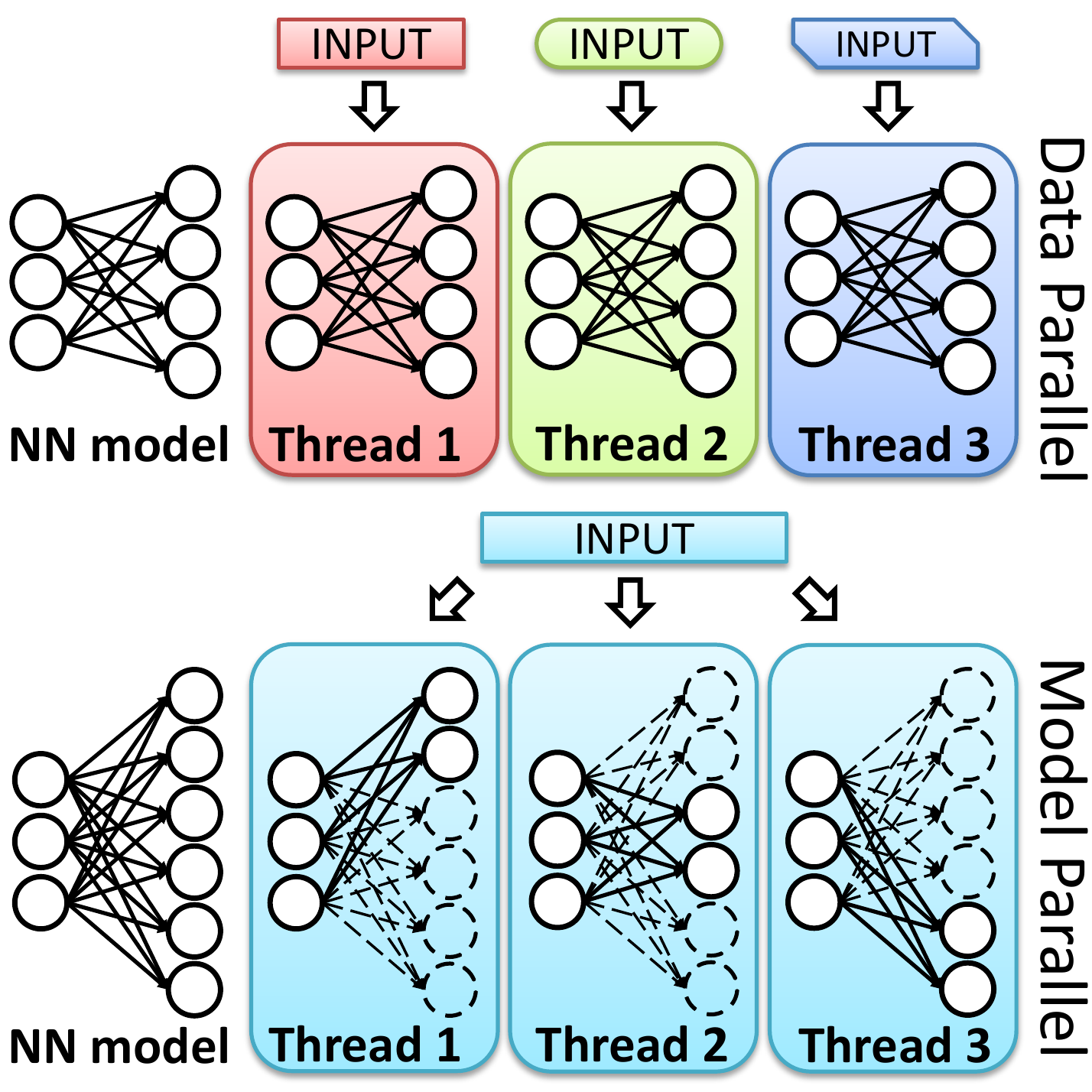}
        \caption{\tonic-NFP data parallel and model parallel modes to process NNs.}
        \label{fig:data_vs_model_parallel}
\end{figure}

\begin{figure}[t!]
        \centering
        \includegraphics[width=1\columnwidth]{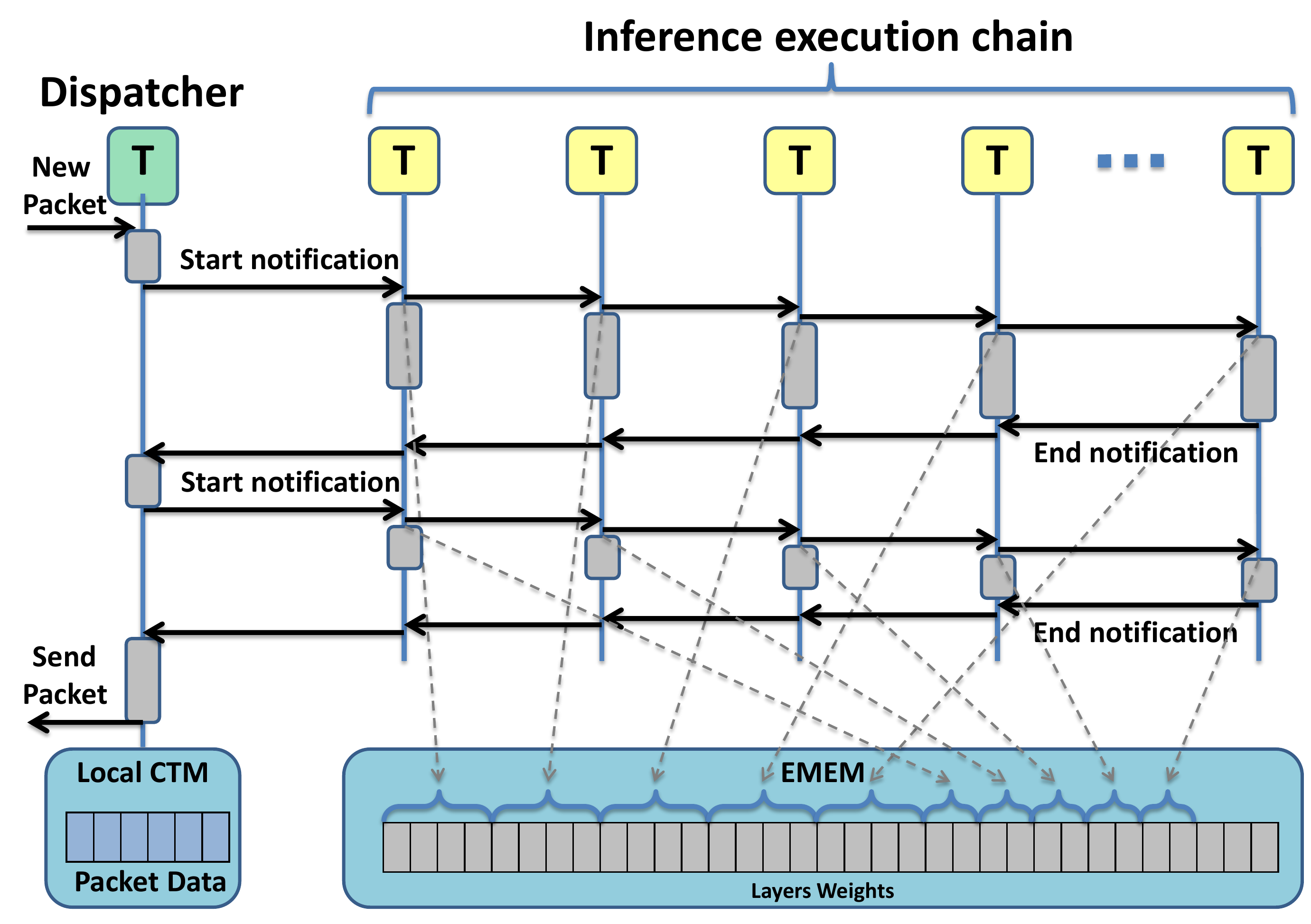}
        \caption{MEs' threads dedicated to inference are organized in a notification chain that is executed mostly in parallel. All threads read neuron's weights from a contiguous memory area located in the EMEM. The network input is encapsulated in the packet data, located in the CTM of the dispatching thread's island.}
        \label{fig:tserver}
\end{figure}

When NNs are large, running them in a single ME's thread would take a long time, making 
the use of multiple threads more effective, even if synchronization among threads incurs 
some overhead. Furthermore, the use of multiple threads allows the NFP to context switch 
them when waiting for memory, hiding memory access time and enabling a more effective use 
of the MEs' processing power. For this reason depending on the NN size, we provided two 
different operation modes for \tonic over Netronome: (1) data parallel and (2) model parallel (cf. Fig~\ref{fig:data_vs_model_parallel}). 
The former, described in \S\ref{sec:impl}, is preferable when NN size is small. The latter, 
instead, when NN is big. In Model Parallel mode MEs' threads are configured with two different 
types of processing tasks. A first subset of threads registers for packet reception notification. 
We call such threads \textit{dispatchers}, and distribute them evenly across the available 
MEs and islands. In our implementation, we empirically found that two of such threads per ME 
is enough to achieve our performance goals. The second subset of threads, named \textit{executors}, 
is instead dedicated to the processing of NNs, and organized in an \textit{execution chain}. 
At boot time, dispatchers registers themselves for packet reception notifications. When a 
new network packet is received, the notified dispatcher parses the packet and performs the 
regular forwarding function. If a NN processing is triggered, a NN processing function is 
executed by the MEs' threads in the execution chain. The dispatcher works as a coordinator for 
the NN processing, by starting the processing of one NN layer, and waiting for the result 
before starting the processing of the next layer. Figure~\ref{fig:tserver} depicts this process.

\vspace{0.1in}
\noindent \textbf{Execution Chain}
The execution chain is statically defined, with each thread knowing its predecessor and
successor threads at boot time. To start processing a layer, the dispatcher notifies the
first thread in the chain with a \textit{start notification}, which is then propagated
throughout the chain, with each thread notifying its successor, until the last thread.
After receiving the start notification, and sending it to the next thread in the chain,
a thread performs the computation of a subset of the current layer's neurons, with the
actual number depending on the number of neurons in the layer and the level of configured
execution parallelism. Each of the threads is an executor from the perspective of \tonic,
so the configured number of executors (threads) defines the level of parallelism.
For instance, for a layer with 4096 neurons, and using 128 executors, each executor would
be in charge of computing 32 neurons. This effectively means, e.g., that \tonic would use
32 MEs, and 4 threads per ME, with each of the threads processing 32 neurons: $32 \times 4 \times 32 = 4096$.
The execution of the neurons happens like in the data parallel case, with the main difference
being that the model's weights are stored in the DRAM-backed EMEM. Here, the weights are placed
in a contiguous memory space, which allows an executor to directly point to the weights of
a set of neurons given its position in the execution chain.
At the end of a layer computation, each executor writes its final result to the global IMEM
memory, from which the dispatching thread can read it.
The last executor in the chain sends an \textit{end notification} to its predecessor after
writing its portion of the result to IMEM. The notification is propagated backward through
the chain as all executors conclude their computations and write their results, until it is
received by the dispatcher. Here, either the computation for a new layer is started or the
final result is collected.

It is worth noticing that the spatial organization of the MEs, which are
distributed across islands and at different distances from each other, makes more efficient
the notification chain, when compared to other mechanisms, such as broadcast messages. That
is, the notification system is faster even if some MEs remain idle while waiting to propagate
the end notification back to the dispatcher. Unfortunately, explicit notifications are required
since each ME's execution time depends on the memory reads from the EMEM, which may take a
variable amount of time. Second, the notification propagation time is relatively short, making
the use of an asymmetric number of threads (or neurons to compute) per-ME inefficient. For
instance one could assign more work to MEs early in the chain, but this in fact rises a problem
of stragglers that harms the overall performance.

\begin{figure*}[!h]
	\centering
	\begin{tabularx}{\linewidth}{XXX}
		\includegraphics[width=\linewidth]{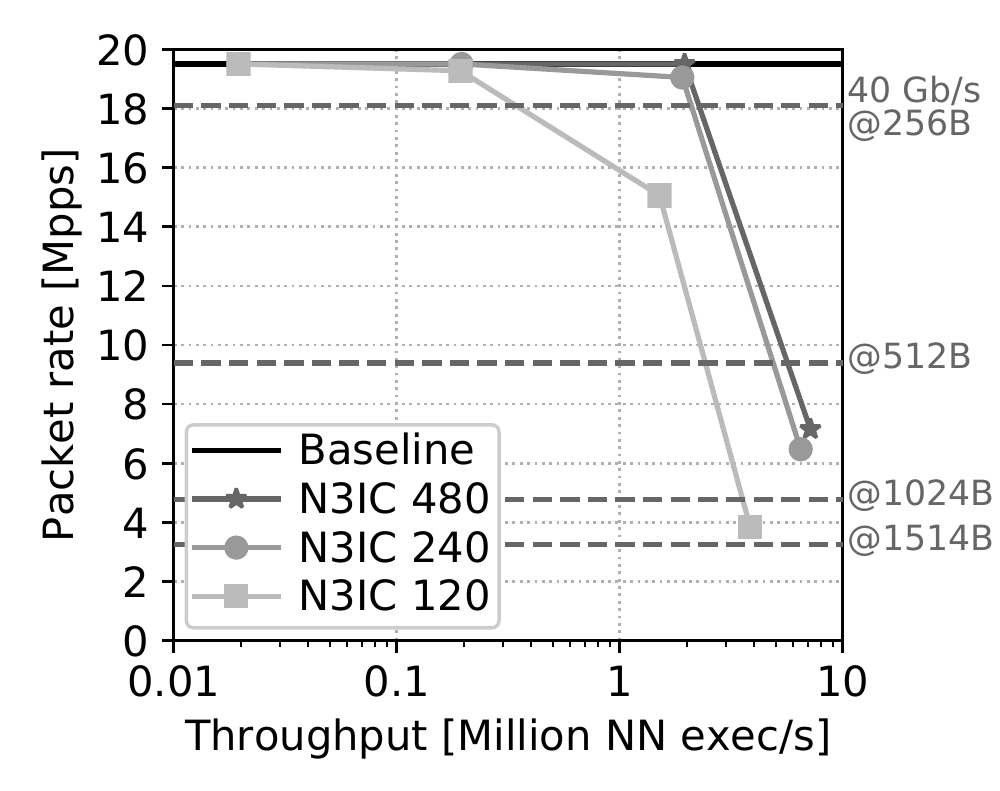}
		\caption{\tonic-NFP data parallel forwarding performance (y axis), when processing 40Gb/s@256B, as a function of the number of flows to analyze (x axis). \emph{X axis in log scale}.}
		\label{fig:data_workload}
		&
		\includegraphics[width=\linewidth]{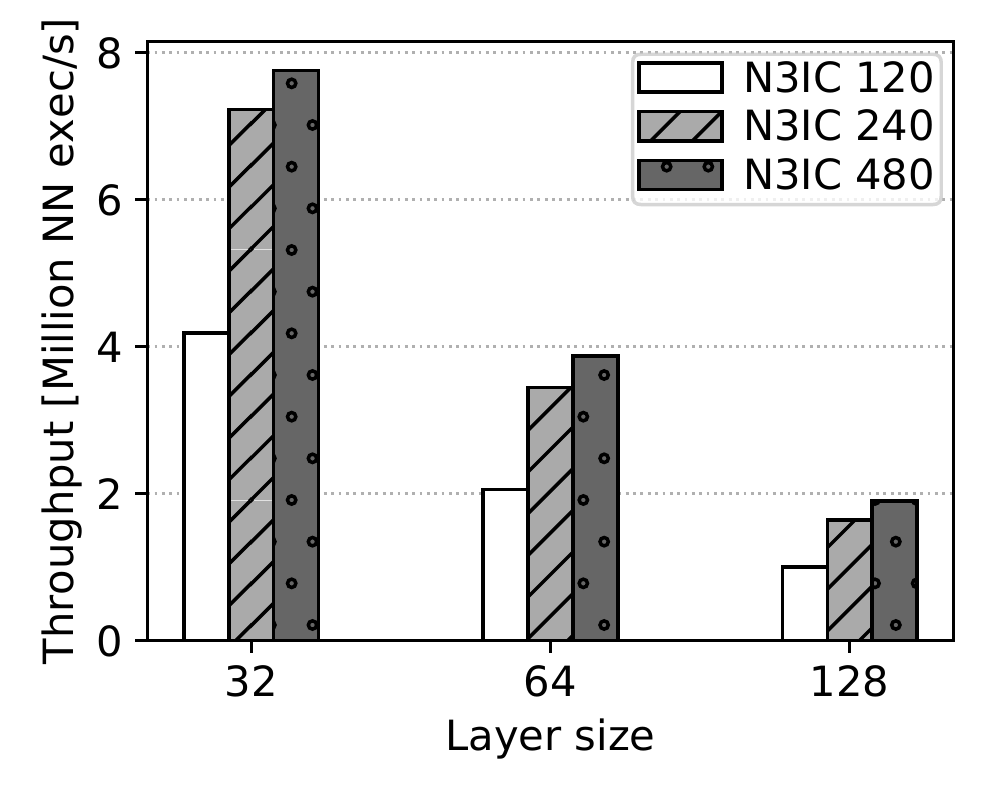}
		\caption{\tonic-NFP data parallel maximum BNN execution throughput (y axis) as a function of the BNN size (x axis). The throughput scales linearly with the BNN size.}
		\label{fig:data_parallel_size}
		&
		\includegraphics[width=\linewidth]{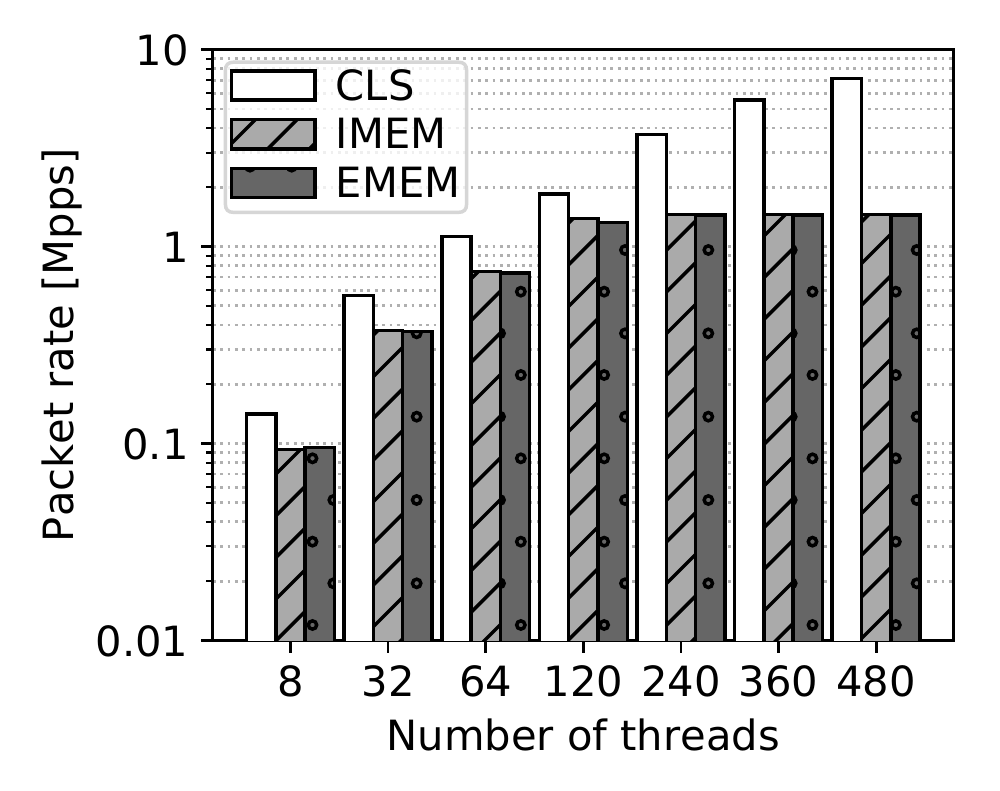}
		\caption{\tonic-NFP data parallel maximum BNN execution throughput (y axis) as a function of the number of used threads (x axis), and for different memories. \emph{Y axis in log scale}.}
		\label{fig:memory_mb}
	\end{tabularx}
	\vspace{-0.8cm}
\end{figure*}

\begin{figure*}[!h]
	\centering
	\begin{tabularx}{\linewidth}{XXX}
		\includegraphics[width=\linewidth]{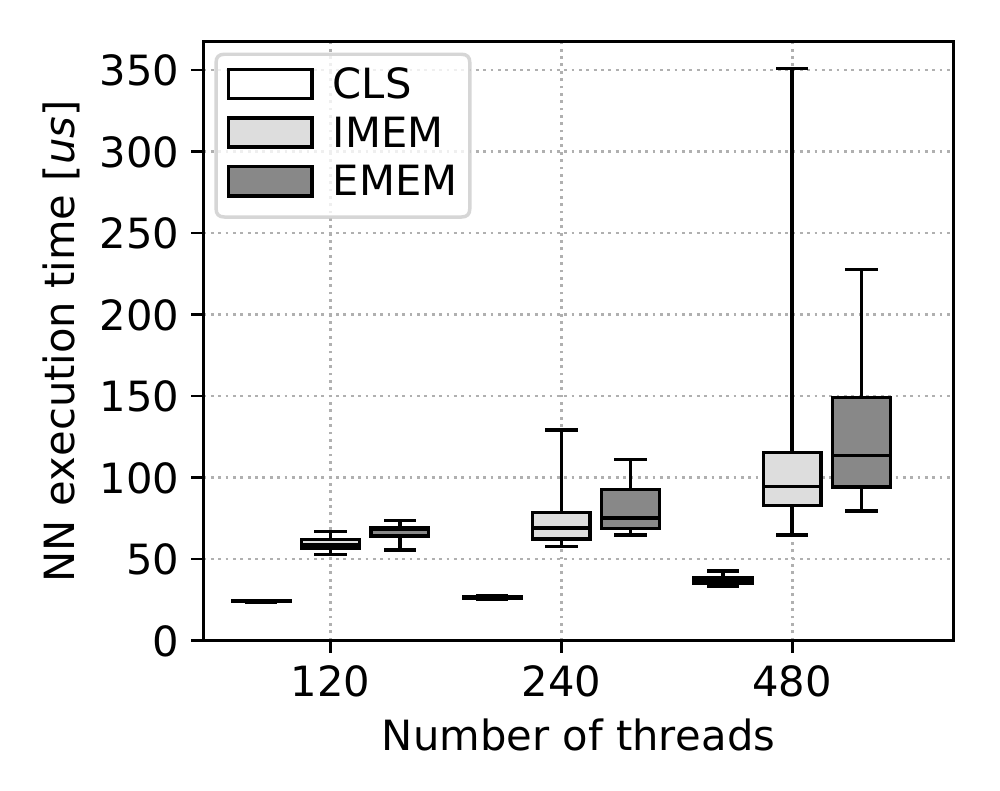}
		\caption{Micro-benchmark of the \tonic-NFP data parallel BNN execution latency (y axis) as a function of the number of threads (x axis), and for different memories.}
		\label{fig:memory_latency}
		&
		\includegraphics[width=\linewidth]{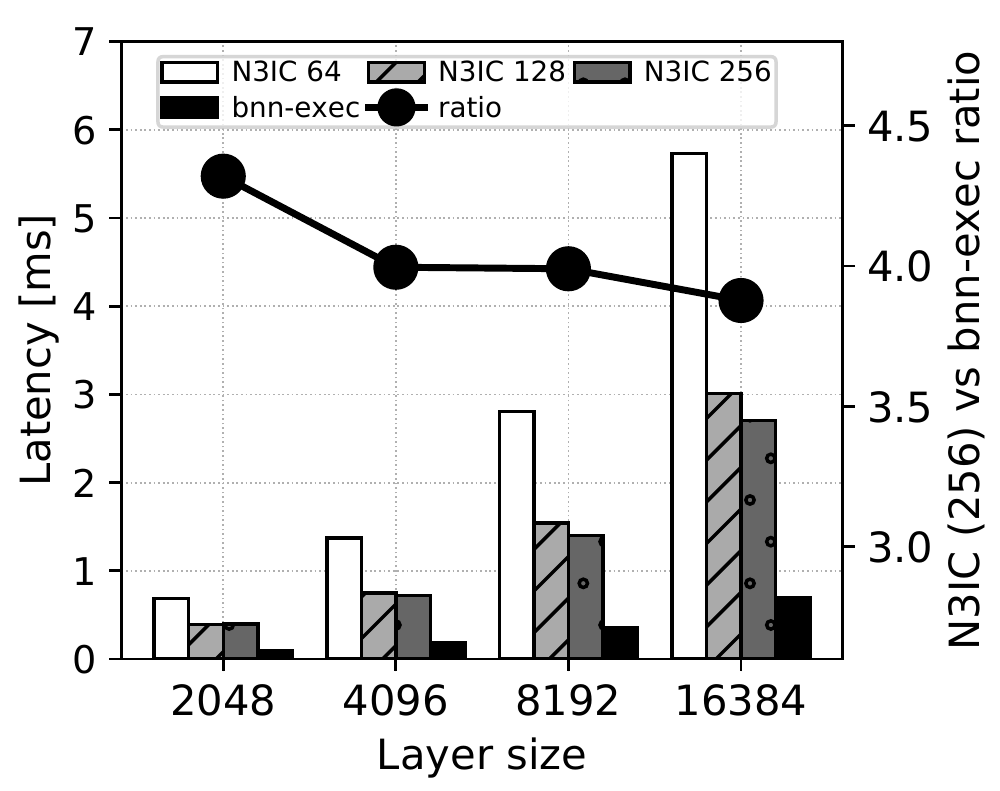}
		\caption{\bnne and \tonic-NFP model parallel processing latency (left y axis) for an FC layer, when varying FC's size (x axis) and number of \tonic-NFP threads. Right y axis shows the ratio between \tonic-NFP with 256 threads and \bnne.}
		\label{fig:latency}
		&
		\includegraphics[width=\linewidth]{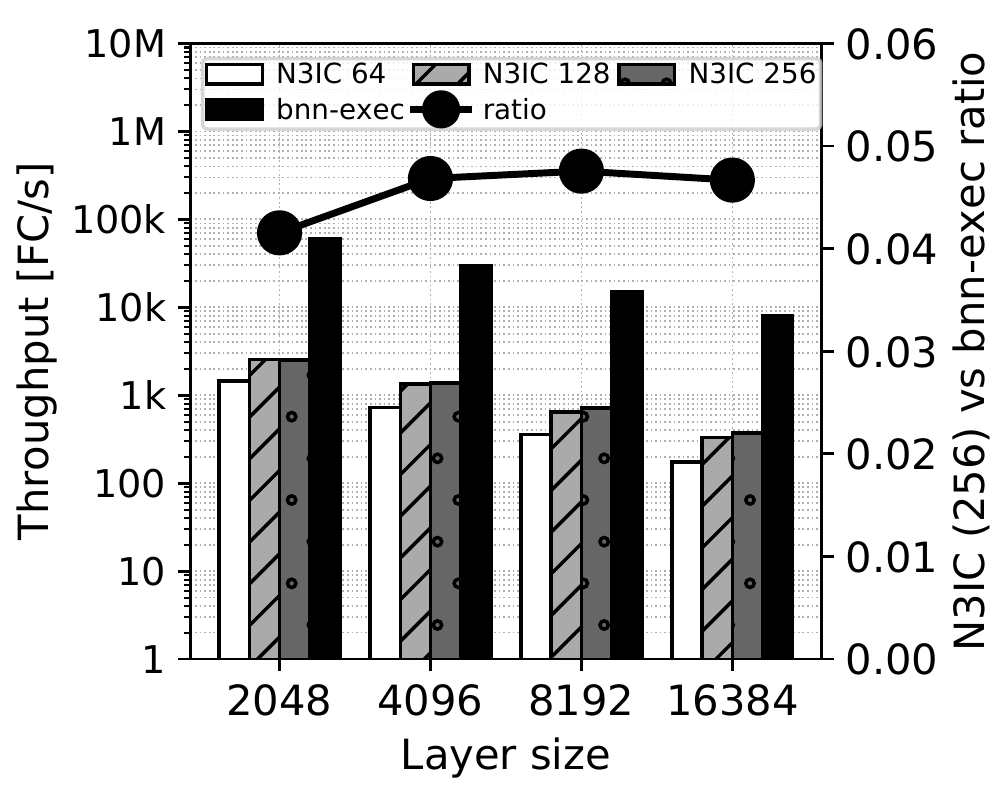}
		\caption{\bnne and \tonic-NFP model parallel throughput (left y axis, \emph{log scale}) for an FC layer, when varying FC's size (x axis) and number of \tonic-NFP threads. Right y axis shows the ratio between \tonic-NFP with 256 threads and \bnne.}
		\label{fig:throughput}
	\end{tabularx}
	\vspace{-0.8cm}
\end{figure*}

\section{Evaluation details}
\label{sec:appendix_netronome_eval_details}
\subsection{NFP4000}
During the development and evaluation of \tonic-NFP we run a number of benchmarks to 
understand the nuances of the NFP4000 architecture, and to tune our implementation. 
In all the tests, the system is always loaded with 40Gb/s@256B. Next we report a subset 
of those results.

\subsubsection{Benchmarks for small NNs (data parallel)}
To better characterize the data parallel performance, we measured the packet forwarding 
capabilities and analyzed flow/s the system achieves when using different flow arrival 
rates, in the orders of 10k, 100k and 1M new flows per second, and executing a NN inference 
for each new received flow. This should cover a wide spectrum of possible workloads~\cite{silkroad}. 
Furthermore, we changed the number of NFP's threads used by \tonic, and measured the NN 
execution time for different configurations. Figure~\ref{fig:data_workload} shows that 
\tonic matches the baseline performance by using 120 threads, i.e., 30 more threads than 
baseline, when handling 200k new flows per second and performing as many NN executions 
per second. This confirms that the computation of the per-flow statistics is a memory-bound 
operation for the NFP4000, which therefore has idle computing resources that could be leveraged 
for NN execution. When further increasing the threads to 240 and 480, \tonic can come close 
to, or match, the baseline performance even while processing about 2M NN executions per 
second, as mentioned in \S\ref{sec:evaluation}.
To check the maximum achievable throughput under heavy load, we configured the NFP to 
process the traffic classification's NN for each received packet, i.e., a stress test. 
In this case, Figure~\ref{fig:data_workload} shows \tonic can 
forward 7.1Mpps, i.e., line rate of 40 Gb/s for packet sizes bigger than 512B, while running 
a NN for each packet using the 480 threads configuration.

\vspace{0.1in}
\noindent\textbf{NN Size (Figure~\ref{fig:data_parallel_size})}. In data parallel mode we 
placed a copy of the the BNN's weights in each of the available CLS memories, since each 
island is provided with a dedicated one. The weights are accessed in read-only mode and 
shared among all the island's threads. We can fit, at most, about 32k weights in CLS\footnote{This 
number is also affected by the number of layers and number of neuron per layers, since each 
thread allocates in CLS the variables required to hold intermediate computation results.}. 
Figure~\ref{fig:data_parallel_size} shows how varying the size of an FC scales linearly 
the \tonic-NFP maximum throughput. The tested layer has 256 inputs, and we run it with a 
different number of neurons: 32 (8.1k weights), 64 (16.3k weights), 128 (32.7k weights).

\vspace{0.1in}
\noindent\textbf{Memory benchmarks}. To understand the impact of the memory selected to 
store NNs' weights, we re-run the stress test using the IMEM and EMEM in place of the CLS, 
and measure both throughput and NN execution latency. \textbf{Figure~\ref{fig:memory_mb}} 
shows that throughput lowers to 1.4Mpps in both cases. Likewise, \textbf{Figure~\ref{fig:memory_latency}} 
shows that the NN execution latency is significantly worse. In particular, the 95-th 
percentile of the latency when using the CLS is 42\si{\micro\second}, instead, the use 
of IMEM and EMEM incurs a much larger variation, with a 95-th percentile inference time 
of 352\si{\micro\second} and 230\si{\micro\second}, respectively. This latency variability 
is due to the need to share the NFP4000's data buses among multiple threads. Interestingly, 
although generally faster, there are cases in which using the IMEM is slower than using 
the EMEM. We believe this is an artefact of the NFP's memory access arbiter.

\subsubsection{Benchmarks for big NNs (model parallel)}
We measured the impact of using a different number of threads and different FC sizes when 
running \tonic-NFP in model parallel mode. We compared the results to those achieved by 
\bnne when running on a single core for latency measurements, and on 4 cores for throughput ones. 
We defined the maximum batch size \bnne can use setting a maximum of 7ms for the processing 
latency~\cite{tpu}. The 7ms execution latency constraint allows \bnne to run with a batch 
size of 64, 32, 16 and 8 for the 2k, 4k, 8k and 16k neurons layers, respectively. The layer 
has 4096 inputs.

\vspace{0.1in}
\noindent\textbf{Latency (Figure~\ref{fig:latency})}.
For layers between 2k and 16k neurons (8M to 67M weights), \tonic-NFP achieves a processing 
latency which is 4 times higher than \bnne's one, varying between 400\si{\micro\second} and 
2700\si{\micro\second}. Considering that the Haswell CPU has a clock frequency more than 4 
times higher than NFP's 800MHz, i.e., each operation is effectively executed in less than a 
fourth of the time, this shows that the NFP is slightly more efficient than the Haswell in 
performing the FC layer operations.

\vspace{0.1in}
\noindent\textbf{Throughput (Figure~\ref{fig:throughput})}.
\tonic-NFP, though unable to perform batching, and using only a subset of the NFP resources, 
can still provide 4-5\% of the \bnne throughput running on a much more powerful Intel CPU. 
Here, take into account that the NFP provides only 3MB of SRAM that have to be shared with 
with packet processing function, while the CPU's over 10MB of L3 cache are dedicated to \bnne 
processing. Furthermore, unlike the CPU (4 cores) that is dedicated to the processing of NNs, 
\tonic-NFP performs such processing while forwarding 18.1Mpps.

\begin{figure}[t!]
	\includegraphics[width=.7\columnwidth]{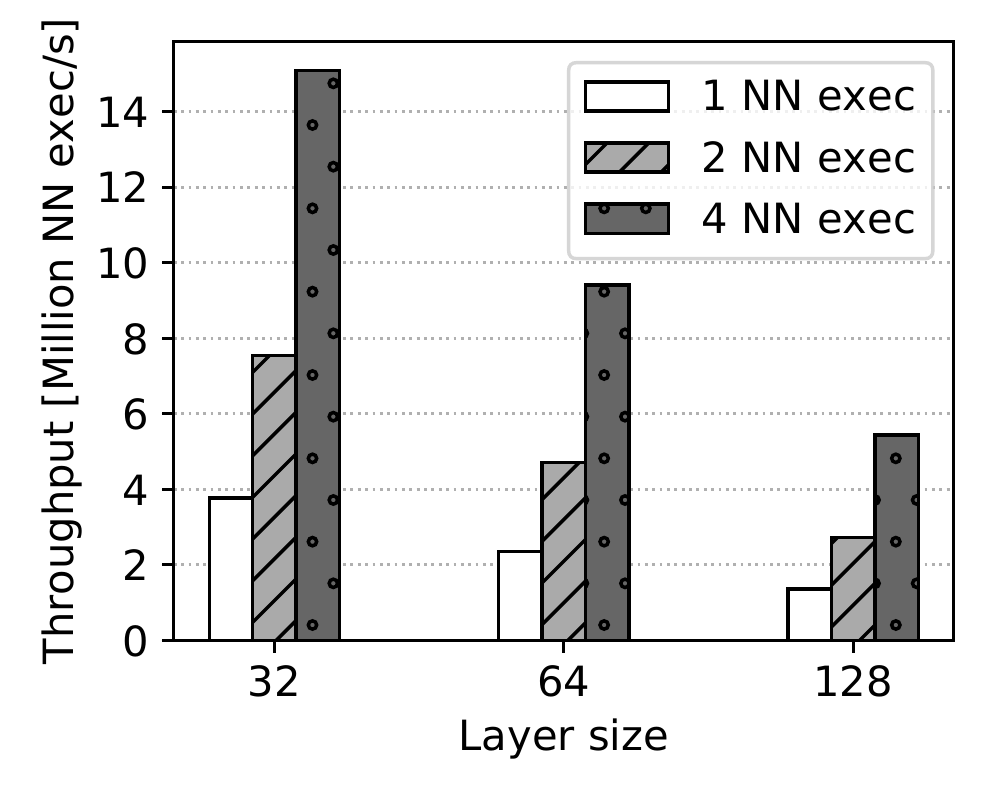}
    \vspace{-0.5cm}
    \caption{\tonic-FPGA throughput when processing different NN sizes and using multiple NN Executor modules. The NN has a single layer with 256b of input.}
    \label{fig:tput_fpga_scaling_per_layer_size}
\end{figure}

\begin{figure}[t!]
	\includegraphics[width=.7\columnwidth]{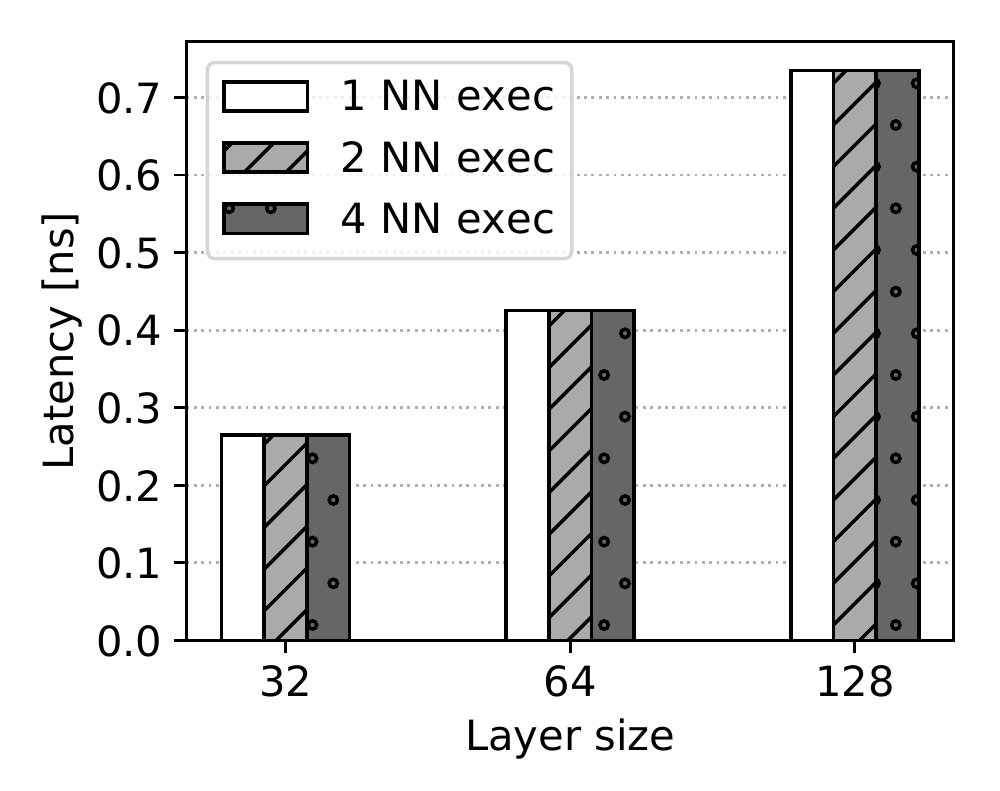}
    \vspace{-0.5cm}
    \caption{\tonic-FPGA latency when processing different NN sizes and using multiple NN Executor modules. The NN has a single layer with 256b of input.}
    \label{fig:latency_fpga_scaling}
\end{figure}

\begin{figure}[t!]
\begin{minipage}[t]{0.32\linewidth}
    \includegraphics[width=\columnwidth]{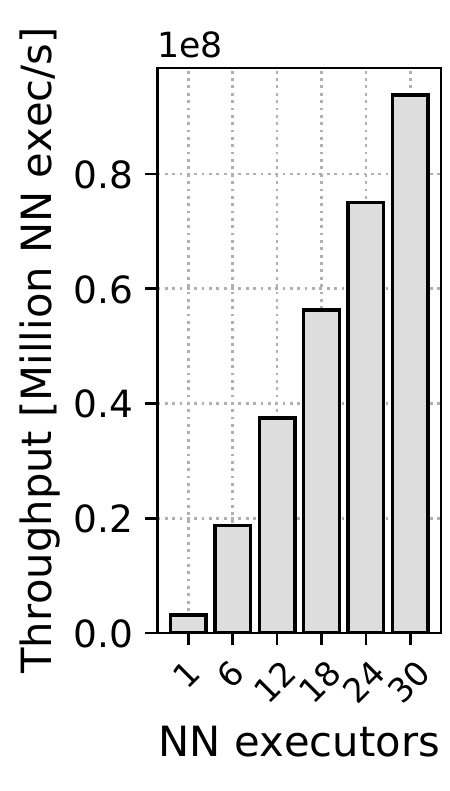}
    \vspace{-0.8cm}
    \caption{\tonic-FPGA TPUT scaling}
    \label{fig:tput_fpga_scaling}
\end{minipage}%
    \hfill%
\begin{minipage}[t]{0.32\linewidth}
    \includegraphics[width=\columnwidth]{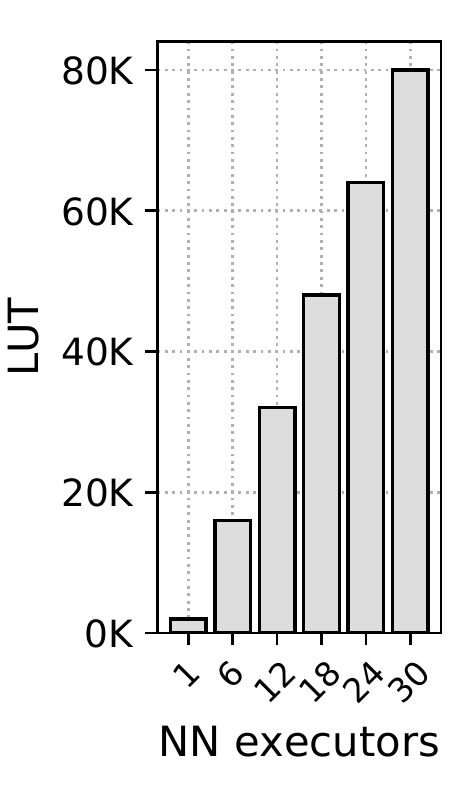}
    \vspace{-0.8cm}
    \caption{\tonic-FPGA LUT scaling}
    \label{fig:lut_fpga_scaling}
\end{minipage} 
\hfill%
\begin{minipage}[t]{0.32\linewidth}
    \includegraphics[width=\columnwidth]{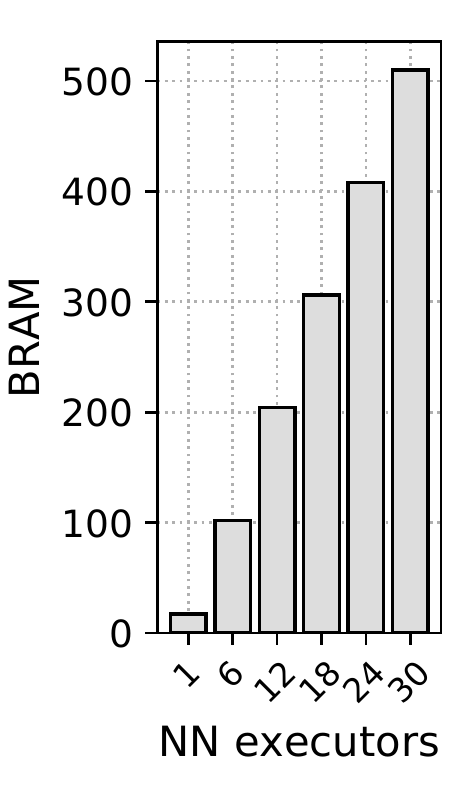}
    \vspace{-0.8cm}
    \caption{\tonic-FPGA BRAM scaling}
    \label{fig:bram_fpga_scaling}
\end{minipage} 
\end{figure}

\subsection{NetFPGA}
We designed the NN Executor module in HDL. This allows us to provide predictable performance, 
i.e., throughput and latency, when performing NN inference.

\vspace{0.1in}
\noindent\textbf{NN Size (Figure~\ref{fig:tput_fpga_scaling_per_layer_size} and Figure~\ref{fig:latency_fpga_scaling})}.
Figure~\ref{fig:tput_fpga_scaling_per_layer_size} shows how varying the size of an FC scales 
linearly the \tonic-FPGA maximum throughput. The tested layer has 256 inputs, and we run it 
with a different number of neurons: 32 (8.1k weights), 64 (16.3k weights), 128 (32.7k weights). 
Increasing the number of NN Executor modules linearly scales the throughput. Since each module 
is dedicated to the execution of a NN, adding more modules does not impact the latency of 
execution, which is only affected by the size of the processed network (cf. Figure~\ref{fig:latency_fpga_scaling}).

\vspace{0.1in}
\noindent\textbf{Resources scaling (Figure~\ref{fig:tput_fpga_scaling}, Figure~\ref{fig:lut_fpga_scaling} and Figure~\ref{fig:bram_fpga_scaling})}.
\tonic-FPGA performance can be increased by deploying multiple NN Executor in parallel. Since 
the logic to manage multiple modules is negligible, also the required FPGA resources scale linearly.
Figure~\ref{fig:tput_fpga_scaling} shows the maximum throughput performance when running the 
anomaly detection NN (cf. \S\ref{sec:use_cases}), during the stress test explained in \S\ref{sec:evaluation}. 
Each NN Executor module increases by about 1.8M inferences per second the obtained performance.
Figure~\ref{fig:lut_fpga_scaling} and Figure~\ref{fig:bram_fpga_scaling} show that the LUTs 
and BRAMs resources also scale linearly with the number of NN Executors. Here, it is worth 
noticing that the use of BRAMs can be considerably optimized. In the current setting, each 
NN Executor has a dedicated CAM element to store the NN weights. However, weights are read-only, 
thus, sharing a CAM module across multiple NN Executors can be achieved with relatively little 
effort. We did not provide such optimizations, since a single NN Executor could already achieve 
the performance goals we set for \tonic-FPGA. 

\section{Use cases details}
We present three use cases to demonstrate the versatility of \tonic. The results are summarized 
in Table~\ref{tab:bnn}.

\subsection{Classification and Anomaly Detection}
We consider two typical networking applications: traffic classification and anomaly detection. 
The former aims at training a NN for application identification and requires the system to be 
able to extract per packet features. The latter focuses on discovering suspicious data flows 
by analyzing flow level features in the network traffic.

\vspace{0.1in}
\noindent\textbf{Datasets}.
For the traffic classification use case, we used UPC-AAU dataset~\cite{bujlow2015independent} 
of more than 750K flows, including per packet traffic traces from 76 commonly used applications 
and services, which account for 55GB of packet-level data\footnote{https://cba.upc.edu/monitoring/traffic-classification}. 
For the anomaly detection use case we used the the UNSW-NB15 dataset~\cite{UNSW_NB15_paper}, 
which provides over 170K network flows labeled in two main categories, i.e., good, bad, and 
their flow-level statistics\footnote{https://www.unsw.adfa.edu.au/unsw-canberra-cyber/cybersecurity/ADFA-NB15-Datasets/}.
In our experiments, we used only the 16 most important features by computing the chi-squared 
value between each feature and the class label~\cite{forman2003extensive}. Furthermore, to 
train a binary classifier, we only use features that can be computed in the NIC hardware. Hence, 
we ignored features related to the packets' content, which we assumed encrypted. For the UPC-AAU 
dataset, we trained a binary classifier to detect encrypted BitTorrent flows.

\begin{figure}[ht!]
        \centering
        \includegraphics[width=.7\columnwidth]{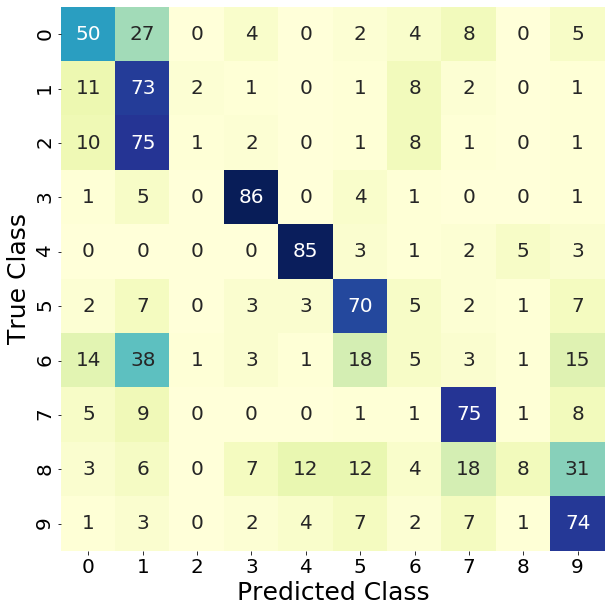}
        \caption{Confusion matrix for multiclass classification on  UPC-AAU dataset. Numbers show the accuracy (\%). Rows and columns shows the class code in Table~\ref{tabel:UPC-AAU}.}
        \label{fig:mcc_upc}
\end{figure}

\vspace{0.1in}
\noindent\textbf{MLP Classifier}. 
For the traffic classification use case, we first trained a regular MLP to classify the traffic 
in 10 classes provided by the UPC-AAU dataset\footnote{Although in UPC-AAU dataset~\cite{bujlow2015independent} 
there 76 different class, there are not enough data samples for each class to train a reliable 
multiclass classifier on it. For this reason, we choose the 10 biggest classes in terms of 
number of available samples.} (see Table~\ref{tabel:UPC-AAU}), achieving 69.4\% accuracy. When 
training a similar binarized MLP model, we achieve 59.1\% accuracy. Figure~\ref{fig:mcc_upc} 
shows the classification confusion matrix for the binarized MLP. As we can see, there are some 
classes, such as code 1,2,6, and 8 in Table~\ref{tabel:UPC-AAU}, which cannot be easily distinguished 
from other classes. Furthermore, we could achieve this classification accuracy only using a larger 
binarized MLP with 256 neurons in each of the two hidden layers. Using the binarized MLP presented 
in \S\ref{sec:use_cases} only achieves about 14\% accuracy. To address the issue, we transformed 
the classification problem in a binary classification. That is, we trained the MLP to classify 
only between BitTorrent traffic and non-BitTorrent traffic. In such a setting we achieved 96.2\% 
accuracy for the regular MLP and 88.6\% for a binarized MLP with only 32, 16, 2 neurons.
This experience shows an interesting application of \tonic, which can be used as a pre-filter 
to classify traffic at a coarse grain. By doing so, it is often possible to shred a significant 
amount of network traffic that does not need further analysis, and dedicate a reduced amount of 
host system's resources to the analysis of network traffic. That is, the host system would need 
to analyze only the subset of traffic that could not be classified by the NIC.

\begin{table}[]
        \begin{tabular}{lll}
                \multicolumn{1}{c}{Code} & \multicolumn{1}{c}{Name}          & \multicolumn{1}{c}{Size} \\\hline
                0                      & bittorrent-all-encrypted          & 25000                                 \\
                1                      & bittorrent-outgoing-non-encrypted & 25000                                 \\
                2                      & emule-outgoing-non-obfuscated     & 10960                                 \\
                3                      & pandomediabooster                 & 12070                                 \\
                4                      & rdp                               & 25000                                 \\
                5                      & web-browser                       & 24914                                 \\
                6                      & dns                               & 10761                                 \\
                7                      & new-samba-session-service         & 25000                                 \\
                8                      & ntp                               & 17834                                 \\
                9                      & ssh                               & 25000\\
                \hline
        \end{tabular}
        \caption{Traffic classes of UPC-AAU with at least 10000 samples.}\label{tabel:UPC-AAU}
\end{table}

We configured an MLP with 3 FCs of 32, 16, 1 neurons for the anomaly detection use case as well.
For both use cases, since each selected feature's numeric value falls in the range [0, 65k], we 
represented them using 16b for each, and provide each bit as separated input to the MLP. With 
this configuration, the MLP has a total of 8.7k weights, and a memory requirement of 35KB, 
assuming 4B to represent each weight. We then performed training using a typical state-of-the-art 
approach, which employs a back-propagation algorithm with Adam optimizer, Dropout on the output 
of each hidden layer with probability 0.25, and binary cross-entropy as loss function. The 
trained classifier achieves 90.3\% accuracy. 

\vspace{0.1in}
\noindent\textbf{Binarized MLP}.
We then designed a binarized MLP to process the same set of features. The binarized MLP has 32, 
16, 2 neurons and a total of 8.7k weights. Being binary weights, the overall required memory 
is only 1KB. Binarization mainly consists in applying a training technique that ensures that 
the weights converge to values included in the range [-1, 1] and normally distributed around 
the 0. In particular, we apply the binarization technique from Courbariaux and Bengio~\cite{trainingBNN}, 
and again train the network using back-propagation, Dropout with probability 0.25, Adam optimizer 
and a squared hinge loss function. The trained MLP's weights obtained after training are still 
real numbers, thus, an additional final step is required to obtain the binarized weights. This 
step is a simple application of a step function to the weights, which are set to 0 if their value 
is lower than 0, or to 1 otherwise. After this, the binarized MLP achieves a 85.3\% and 88.6\% 
accuracy on the test set for UNSW-NB15 and UPC-AAU, respectively. We observe that the BNN provides
only 5 and 8 percentage points lower accuracy than the non-binarized version. 

\begin{table}[!t]
	\caption{Memory requirements and accuracies of the NNs used by the presented use cases.}
	\label{tab:bnn}
	\begin{center}
		\begin{small}
			\begin{sc}
			\begin{tabular}{cccccccc}
			\multicolumn{1}{c||}{\multirow{2}{*}{Data}} & \multicolumn{1}{c||}{\multirow{2}{*}{NN size}} &\multicolumn{2}{c||}{Memory}                             & \multicolumn{2}{c}{Accuracy}                             &  &  \\ \cline{3-6}
			\multicolumn{1}{c||}{}      &\multicolumn{1}{c||}{}      & \multicolumn{1}{c|}{MLP}    & \multicolumn{1}{c||}{BIN}   & \multicolumn{1}{c|}{MLP}    & \multicolumn{1}{c}{BIN}    &  &  \\ \cline{1-6}
			\multicolumn{1}{c||}{UNSW}  & \multicolumn{1}{c||}{32,16,2}  & \multicolumn{1}{c|}{35KB}   & \multicolumn{1}{c||}{1.1KB}   & \multicolumn{1}{c|}{90.3\%} & \multicolumn{1}{c}{85.3\%} &  &  \\
			\multicolumn{1}{c||}{UPC}   & \multicolumn{1}{c||}{32,16,2}  & \multicolumn{1}{c|}{35KB}   & \multicolumn{1}{c||}{1.1KB}   & \multicolumn{1}{c|}{96.2\%} & \multicolumn{1}{c}{88.6\%} &  &  \\
            \multicolumn{1}{c||}{NS3}   & \multicolumn{1}{c||}{32,16,2}  & \multicolumn{1}{c|}{21.6KB}   & \multicolumn{1}{c||}{676B}   & \multicolumn{1}{c|}{92.0\%} & \multicolumn{1}{c}{90.0\%} &  &  \\
            \multicolumn{1}{c||}{NS3}   & \multicolumn{1}{c||}{64,32,2}  & \multicolumn{1}{c|}{47.2KB}   & \multicolumn{1}{c||}{1.5KB}   & \multicolumn{1}{c|}{94.0\%} & \multicolumn{1}{c}{90.0\%} &  &  \\
            \multicolumn{1}{c||}{NS3}   & \multicolumn{1}{c||}{128,64,2}  & \multicolumn{1}{c|}{110.8KB}   & \multicolumn{1}{c||}{3.4KB}   & \multicolumn{1}{c|}{94.0\%} & \multicolumn{1}{c}{92.0\%} &  &  \\ \cline{1-6}
			\end{tabular}
			\end{sc}
		\end{small}
	\end{center}
\end{table}

\subsection{Network Tomography}

\begin{figure}[t!]
\begin{minipage}[t]{0.48\linewidth}
         \includegraphics[width=\linewidth]{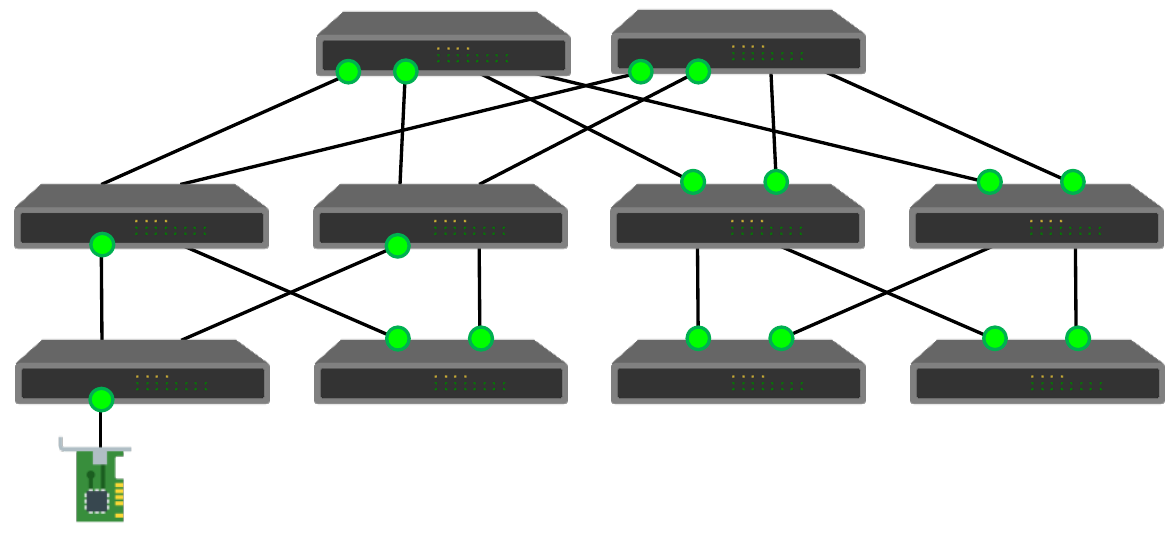}
    \vspace{-0.8cm}
         \caption{ns3 topology for the network tomography use case.}
         \label{fig:simon_appendix}
    \vspace{-0.3cm}
\end{minipage}%
    \hfill%
\begin{minipage}[t]{0.48\linewidth}
         \includegraphics[width=\linewidth]{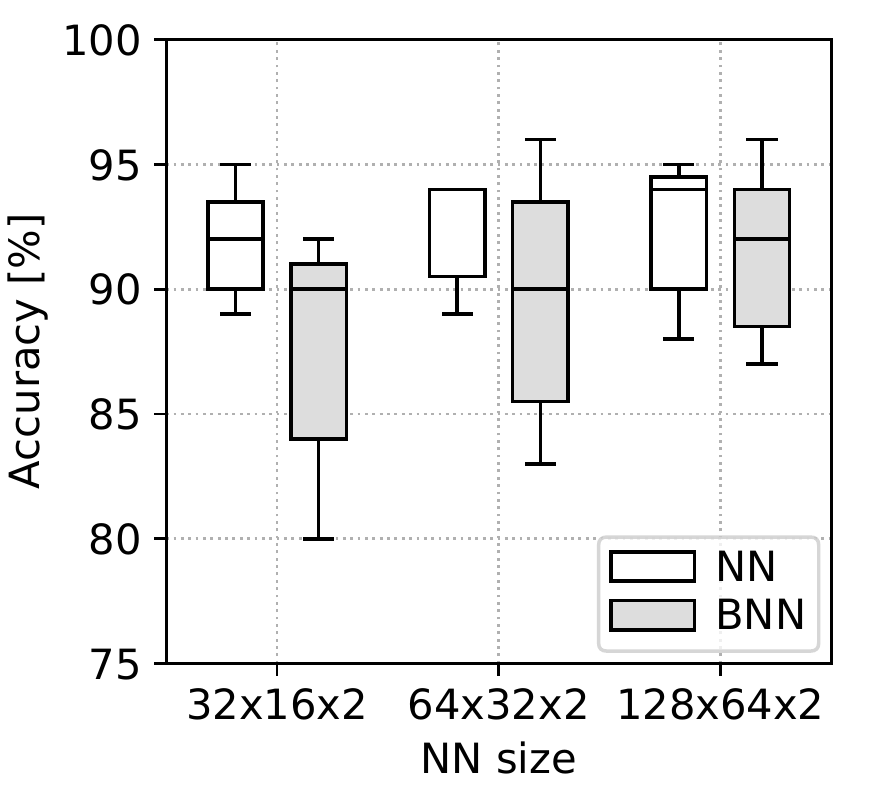}
    \vspace{-0.8cm}
         \caption{Box plot of the accuracies for the predicted queues in the network tomography use case.}
         \label{fig:simon_accuracy_nn_vs_bnn}
    \vspace{-0.3cm}
\end{minipage}
\end{figure}

In this use case, we run SIMON~\cite{geng19}, a network tomography application which infers congestion points and queue sizes in a datacenter network starting from host-to-host network paths' delays.
All the end-hosts periodically exchange probe packets among them and send the measured delays to a centralized node which is then responsible for running the reconstruction engine and whose algorithm has been approximated using a Neural Network.
We implemented a modified version of SIMON where we are interested in quickly detecting the congestion point without estimating the exact size of the queues.
Rather than running a single NN in the centralized node, a set of NICs is in charge of computing the congestion status of the queues.

\vspace{0.1in}
\noindent\textbf{Dataset}.
We simulated a small CLOS-like Fat Tree datacenter network in ns3~\cite{ns3}.
The network, reported in Fig.~\ref{fig:simon_appendix}, includes 10 switches and 32 hosts organized in two pods (4 ToR switches, 4 aggregation switches and 2 core switches).
The datacenter operates under an incast traffic load as described in~\cite{geng19}.
In addition to the traffic, all the servers (except the first one) periodically send a probe packet (once every 10ms) towards the first server to measure the network paths delays.\footnote{For the sake of simplicity we focus on a scenario where SIMON is just run on a single NIC, specifically in the first server, but in the complete scenario every server will probe all the other servers so that multiple NICs, connected under different racks, cover the whole set of queues in the network.}
From any server there are up to 8 distinct paths towards the first server, traversing in total 17 distinct output queues (reported as green dots in Fig.~\ref{fig:simon_appendix}).
Our task is to train multiple NNs, one for each queue, each one in charge of detecting the queue congestion status.

We selected a subset of 19 out of 31 probes in order to keep 1 probe per distinct path.
Our dataset consists of 30k samples, one per each 10ms interval, with 19 features (path delays, in ms) and 17 corresponding outputs (queue sizes, in packets).
We considered 17 independent binary classification problems where the output class is 1 if in a given 10ms interval the corresponding queue is above a configurable threshold, 0 otherwise.

\vspace{0.1in}
\noindent\textbf{MLP Classifier}. 
We first trained a regular MLP, with the same hyperparameters used in the previous use cases, with three different architectures (32x16x1, 64x32x1 and 128x64x1).
The 19 inputs of the MLP are represented using 8 bits and, as in the previous use cases, we provide each bit as separated input to the MLP.
Table~\ref{tab:bnn} reports the memory requirements for the different NN architectures.
The resulting median accuracies range from 92\% to 94\% for an increasing NN size.


\vspace{0.1in}
\noindent\textbf{Binarized MLP}.
We then designed a binarized MLP to process the same set of features with three different architectures (32x16x2, 64x32x2 and 128x64x2), all with 19 inputs and two output neurons.
The drop in the median accuracy when moving from a full precision NN to a binarized NN ranges from 2\% to 4\%.
Fig.~\ref{fig:simon_accuracy_nn_vs_bnn} reports more in detail the distribution of the median accuracies for different NN sizes.

\end{appendix}

\end{document}